\documentclass[10pt,journal,twoside,final]{IEEEtran}

\IEEEoverridecommandlockouts

\makeatletter
\newcommand{\biggg}[1]{{\hbox{$\left#1\vbox to 20.5pt{}\right.\n@space$}}}
\newcommand{\Biggg}[1]{{\hbox{$\left#1\vbox to 23.5pt{}\right.\n@space$}}}
\newcommand{\bigggg}[1]{{\hbox{$\left#1\vbox to 26.5pt{}\right.\n@space$}}}
\newcommand{\Bigggg}[1]{{\hbox{$\left#1\vbox to 29.5pt{}\right.\n@space$}}}
\newcommand{\biggggg}[1]{{\hbox{$\left#1\vbox to 32.5pt{}\right.\n@space$}}}
\newcommand{\Biggggg}[1]{{\hbox{$\left#1\vbox to 35.5pt{}\right.\n@space$}}}
\newcommand{\bigggggg}[1]{{\hbox{$\left#1\vbox to 38.5pt{}\right.\n@space$}}}
\newcommand{\Bigggggg}[1]{{\hbox{$\left#1\vbox to 41.5pt{}\right.\n@space$}}}
\makeatother

\usepackage{amsfonts}
\usepackage{bm,cite,float,amssymb}
\usepackage{subfigure,amsfonts}
\usepackage{mdwmath}
\usepackage{mdwtab}
\usepackage{xcolor,color}
\usepackage{cool}
\usepackage{graphicx}
\usepackage{multicol}
\usepackage{amsmath,lipsum}
\usepackage{balance}
\usepackage{diagbox}
\usepackage{bm}
\usepackage{cuted}
\usepackage{stfloats}
\usepackage{algorithm}%
\usepackage{algorithmicx}
\usepackage{multirow}%
\usepackage{amsthm}%
\usepackage{textcomp}%
\usepackage{manyfoot}%
\usepackage{booktabs}%
\usepackage{algpseudocode}

\graphicspath{{imgs/}}   %%%

\makeatletter
\renewcommand\paragraph{\@startsection{paragraph}{4}{\z@}%
            {-2.5ex\@plus -1ex \@minus -.25ex}%
            {1.25ex \@plus .25ex}%
            {\normalfont\normalsize\itshape}}
\makeatother
\usepackage{epstopdf}

\usepackage[style=0]{mdframed}
\usepackage{showexpl}

\usepackage{cite}

\mdfdefinestyle{exampledefault}{%
rightline=true,innerleftmargin=10,innerrightmargin=10,
frametitlerule=true,frametitlerulecolor=green,
frametitlebackgroundcolor=yellow,
frametitlerulewidth=2pt}

\allowdisplaybreaks

\definecolor{myGreen}{rgb}{0.76, 0.93, 0.63}
\definecolor{myGreen2}{rgb}{0.92, 0.92, 0.92}
\begin{document}

\title{{Dual-UAV-Enabled Secure Communication\\ and Sensing for A2G-ISAC Systems with Maneuverable Jamming}}

\author{Libiao Lou, 
Yuan Liu, 
Fotis Foukalas,  \IEEEmembership{Senior Member,~IEEE,}
Hongjiang Lei,   \IEEEmembership{Senior Member,~IEEE,}
Gaofeng Pan,    \IEEEmembership{Senior Member,~IEEE,}
Theodoros A. Tsiftsis,  \IEEEmembership{Senior Member,~IEEE,}
and Hongwu~Liu,  \IEEEmembership{Senior Member,~IEEE}
\thanks{L. Lou, Y. Liu, and H. Liu are with the School of Information Science and Electrical Engineering, Shandong Jiaotong University, Jinan 250357, China (e-mail: 205029@sdjtu.edu.cn, 240041@sdjtu.edu.cn,  liuhongwu@sdjtu.edu.cn).}
\thanks{F. Foukalas and T. A. Tsiftsis are with the Department of Informatics and Telecommunications, University of Thessaly, 35100 Lamia, Greece (e-mail: foukalas@ieee.org, tsiftsis@uth.gr).}
\thanks{H. Lei is with the Chongqing Key Laboratory of Mobile Communications Technology, Chongqing University of Posts and Telecommunications,
 Chongqing 400065, China (e-mail: leihj@cqupt.edu.cn)}
\thanks{G. Pan is with  the School of Cyberspace Science and Tech
nology, Beijing Institute of Technology, Beijing 100081, China (e-mail:
 gaofeng.pan.cn@ieee.org).}
}
 
\maketitle
\setcounter{page}{1}
\begin{abstract}
In this paper, we propose a dual-unmanned aerial vehicle (UAV)-enabled secure communication and sensing (SCS) scheme for an air-to-ground integrated sensing and communication (ISAC) system, in which a dual-functional source UAV and jamming UAV collaborate to enhance both the secure communication and target sensing performance. From a perspective of hybrid monostatitc-bistatic radar, the jamming UAV maneuvers to aid the source UAV to detect multiple ground targets by emitting artificial noise, meanwhile interfering with the ground eavesdropper. Residual interference is considered to reflect the effects of imperfect successive interference cancellation (SIC) on the receive signal-plus-interference-to-noise ratios, which results in a degraded system performance. 
To maximize the average secrecy rate (ASR), the dual-UAV trajectory and dual-UAV beamforming are jointly optimized subject to the transmit power budget, UAV maneuvering constraint, and sensing requirements. To tackle the highly complicated non-convex ASR maximization problem, the dual-UAV trajectory and dual-UAV beamforming are optimized for the secure communication (SC) purpose and the SCS purpose, sequentially. In the SC phase, a block coordinate descent algorithm is proposed to optimize the dual-UAV trajectory and dual-UAV beamforming iteratively, using the trust-region successive convex approximation (SCA) and semidefinite relaxation (SDR) techniques. Then, a weighted distance minimization problem is formulated to determine the dual-UAV maneuvering positions suitable for the SCS purpose, which is solved by a heuristic greedy algorithm, followed by the joint optimization of source beamforming and jamming beamforming. The simulation results show that the proposed scheme achieves the highest ASR performance among existing benchmark schemes, while the sensing performance is guaranteed by exploiting the maneuverable jamming. The effects of the SIC residual interference on the secure communication and target sensing performance are also verified by the simulation results.     
\end{abstract}


\section{Introduction} 

Integration of sensing and communication (ISAC) is essential for sixth generation (6G) wireless networks to enable extremely ultra-reliable and extremely low-latency connectivity while supporting real-time environmental perception and data transmission\cite{ISAC_Dual-Functional_Wireless_Networks,ISAC_Resource_Allocation}. By unifying both communication and radar sensing functionalities, ISAC improves spectral efficiency, realizes environmental awareness, and paves the way for diverse intelligent applications like autonomous systems, immersive digital-physical interactions, and smart city infrastructures \cite{ ISAC_Waveform_Design,ISAC_VCN,ISAC_Advances_Challenges}. 
Unmanned aerial vehicles (UAVs), which emerged as transformative enablers for air-to-ground (A2G)-ISAC systems, have garnered significant attention from both academia and industry due to their dynamic network deployment, high-resolution sensing, and reliable line-of-sight (LoS) transmissions \cite{UAV_ISAC_EE, beamforming_design, UAV_ISAC_Deployment_and_Precoder,  RIS_UAV_ISAC}.  
Enabled by UAVs, A2G-ISAC facilitates applications across critical domains including disaster response, precision agriculture, smart city monitoring, and autonomous vehicle navigation, leveraging UAVs' maneuverability and adaptive communication-sensing synergy \cite{UAV_ISAC_Sky, UAV_ISAC_IoT, UAV_ISAC_JBT,ISAC_UAV_JCT}.

Due to the open nature of wireless communications and the inherent vulnerabilities of LoS propagation, A2G-ISAC faces growing security challenges, including the exposure of sensitive data to eavesdropping, spoofing, and interference. Constrained by limited onboard computational capabilities and dynamic networking, traditional encryption methods are hard to ensure information security in A2G-ISAC systems. 
Physical Layer Security (PLS) technology, characterized by its minimal reliance on system computational capabilities, constitutes a pivotal mechanism for safeguarding A2G-ISAC's information security. 
In dynamic A2G-ISAC environments, PLS techniques, such as beamforming, 
artificial noise (AN) jamming, trajectory design, and resource allocations, have been applied to mitigate eavesdropping and ensure secure wireless transmissions 
\cite{Secrecy_AAV_ISAC,PLS_ISAC_AAV, UAVs_meet_ISAC_Secure,Security_ISAC_IRS_UAV,Secure_IRS_UAV_ISAC,IRS_UAV_Secure_ISAC}. In \cite{Secrecy_AAV_ISAC}, a UAV was deployed as an aerial dual-functional base-station (BS) to serve ground users and detect eavesdroppers simultaneously. Furthermore, a joint UAV trajectory and beamforming optimization method was proposed to maximize the average secrecy rate (ASR) while ensuring sensing accuracy with respect to untrusted targets in A2G-ISAC systems\cite{PLS_ISAC_AAV}. To achieve the high-precision tracking and maximize the secrecy rate, a real-time UAV trajectory optimization method utilizing the extended Kalman filtering and coordinated beamforming was proposed in \cite{UAVs_meet_ISAC_Secure}. 
With the advancements of intelligent reflecting surface (IRS), several IRS-aided PLS schemes were proposed for A2G-ISAC systems to enhance both the security and sensing performance by jointly optimizing the UAV trajectory and beamforming \cite{Security_ISAC_IRS_UAV,Secure_IRS_UAV_ISAC, IRS_UAV_Secure_ISAC}.  

Nevertheless, most researches on the PLS-based secure communications for A2G-ISAC systems remain confined to single-UAV scenarios, failing to fully exploit the performance gains enabled by multi-UAV coordination. Specifically, existing approaches predominantly assume static or predefined flight trajectories,  overlooking the inherent spatial diversity and cooperative beamforming capabilities offered by distributed UAVs \cite{Multi_UAV_Ssecure}. 
Through coverage extension, interference suppression, and resource scheduling, 
multi-UAV coordination effectively improved system gains for A2G-ISAC \cite{Multi_UAV_ISAC_GCWkshps, Multi_UAV_ISAC_WCSP, Multi_UAV_ISAC_IoTJ, ISAC_UAV_Swarms}. In an A2G-ISAC system consisting of multiple dual-functional UAVs, the collaborative signal processing, UAVs' placement, and power control were jointly explored to enhance the detection performance of the ground target \cite{Multi_UAV_ISAC_GCWkshps}. In addition, the beamforming and UAVs' placement were considered to maximize the minimum detection probability over the target, while ensuring the minimum signal-to-interference-plus-noise ratio (SINR) requirement among ground users \cite{Multi_UAV_ISAC_WCSP}. When multiple dual-functional UAVs were deployed to communicate with multiple ground users and detect a ground target, the user association, beamforming, and UAV trajectory were jointly designed to maximize the weighted sum-rate,  while ensuring the sensing beampattern gain of the target \cite{Multi_UAV_ISAC_IoTJ}.

On the other hand, jamming techniques, including friendly and malicious jamming, were strategically employed in A2G-ISAC systems to degrade the eavesdropping reception quality or legitimate communication integrity, respectively \cite{UAV_ISAC_Malicious_Jamming,UAV_ISAC_Jamming_Attack,UAV_ISAC_PLS_Multiple_Eavesdroppers,UAV_secure_JTR}. 
To mitigate adversarial impacts on the received SINRs, both friendly and malicious jamming necessitate robust beamforming and power allocation schemes in dynamic A2G-ISAC environments. 
By reducing the received SINRs for authorized users, malicious jamming disrupts legitimate communication links, thus degrading PLS performance in A2G-ISAC systems \cite{UAV_ISAC_Malicious_Jamming,UAV_ISAC_Jamming_Attack}.
Conversely, friendly jamming enhanced the PLS performance of A2G-ISAC systems by intentionally introducing controlled interference to obscure eavesdroppers, thereby improving secrecy rate and reducing information leakage risks\cite{UAV_ISAC_PLS_Multiple_Eavesdroppers,UAV_secure_JTR}. 
In \cite{UAV_ISAC_PLS_Multiple_Eavesdroppers}, a stationary jamming UAV is deployed at the geometric center of multiple eavesdroppers to interfere  eavesdropping links, addressing PLS challenges in A2G-ISAC scenarios. Furthermore, to suppress eavesdropping, a maneuverable jamming UAV tracked the moving targets following a straight-line trajectory, while the source UAV dynamically adjusted its trajectory and optimized resource allocation \cite{UAV_secure_JTR}. However, current research on cooperative jamming in A2G-ISAC systems lacks comprehensive consideration of jamming UAV mobility in PLS designs. This oversight restricts the potential for further enhancing PLS performance through dynamic trajectory optimization of jamming UAVs. 

Motivated by the aforementioned works, we propose a dual-UAV-enanbled secure communication and sensing (SCS) scheme for an A2G-ISAC system to enhance both the sensing and  secure communication performance. In particular, the jamming UAV is deployed to kill two birds with one stone, i.e., the jamming UAV transmits the AN signals not only to degrade the eavesdropping reception quality but also to improve the sensing SINR, by formulating a hybrid monostatic-bistatic radar system consisting of the source and jamming UAVs. Through the joint optimization of the dual-UAV trajectory and dual-UAV beamforming, both the secure communication and target sensing performance can be improved significantly. Our main contributions can be summarized as follows:  

\begin{itemize}
\item We propose a dual-UAV-enabled SCS scheme for the A2G-ISAC system, in which the source and jamming UAVs cooperate to carry out dual-functional sensing and secure communication. In particular, the AN signals transmitted by the jamming UAV are utilized to degrade the quality of the eavesdropping link, while the source and jamming UAVs formulate a hybird monostatic-bistatic radar system to sense multiple ground targets. achieving simultaneous secure communication and target sensing optimization.
\item To maximize the ASR, the dual-UAV trajectory and dual-UAV beamforming are jointly optimnized, subject to the transmit power budgets, UAV maneuvering constraints, and sensing requirements. Considering imperfect successive interference cancellation (SIC) at the communication and radar receivers,  residual interference effects on the received SINRs are modeled. To tackle the highly complex non-convex ASR maximization problem, the decoupled the dual-UAV trajectory and dual-UAV beamforming subproblems are solved interatively in a block coordinate descent (BCD) way, using the trust-region successive convex approximation (SCA) and semidefinite relaxation (SDR).
\item  To facilitate determining the dual-UAV locations suitable for dual-functional SCS, the dual-UAV trajectory is first optimized for the secure communication (SC) purpose only; Then, a greedy algorithm is proposed to solve a corresponding weighted-distance minimization problem for the SCS purpose. The simulation results demonstrate that the proposed secure communication scheme obtains the superior secure communication and target sensing performance over the existing benchmark schemes. The residual interference effects on the dual-functional system performance are also revealed.   
\end{itemize}

The remainder of the paper is organized as follows. Section II establishes the system model of the considered A2G-ISAC system and formulates the ASR maximization problem. Sections III and IV present the sub-optimal solutions for the SC and SCS purposes, respectively. Section V clarify the system performance via simulations, while Section VI concludes the work.
 
$Notations$: The statistical expectation is denoted as $\mathbb{E}(\cdot)$. Vectors and matrices are denoted using bold lowercase and uppercase italic letters, respectively. $(\cdot)^*$, $(\cdot)^T$, and $(\cdot)^H$ denote the conjugate, transpose, and Hermitian transpose operations, respectively.  $\bm{0}$ and $\bm{I}$ stand for the all-zero matrix and identity matrix, respectively. $\bm{A} \succeq 0$ represents that 
matrix $\bm A$ is positive semidefinite. $[\bm{A}]_{x,y}$ denotes the element in the $x$-th row and $y$-th column of  matrix $\bm A$. The magnitude and phase of $[\bm{A}]_{x,y}$ are denoted as $\big\vert[{\bm A}]_{x,y}\big\vert$ 
and $\theta^{\bm A}_{x,y}$, respectively. ${\rm{rank}}(\cdot)$ and ${\rm{tr}}(\cdot)$ denote the rank and trace of a matrix, respectively. The Euclidean norm, nuclear norm, and spectral norm are denoted by
$\| \cdot \|$, $\| \cdot \|_*$, and $\| \cdot \|_2$, respectively.  $\mathbb{C}^{M \times N}$ denotes the $M \times N$ complex matrices. $\mathcal{C}\mathcal{N}(\nu, \sigma^2)$ is the  circularly symmetric complex Gaussian (CSCG) distribution with mean $\nu$ and variance $\sigma^2$.

\section{System Model}

\begin{figure}[t]
 \begin{center}
    \includegraphics[width=3.2in]{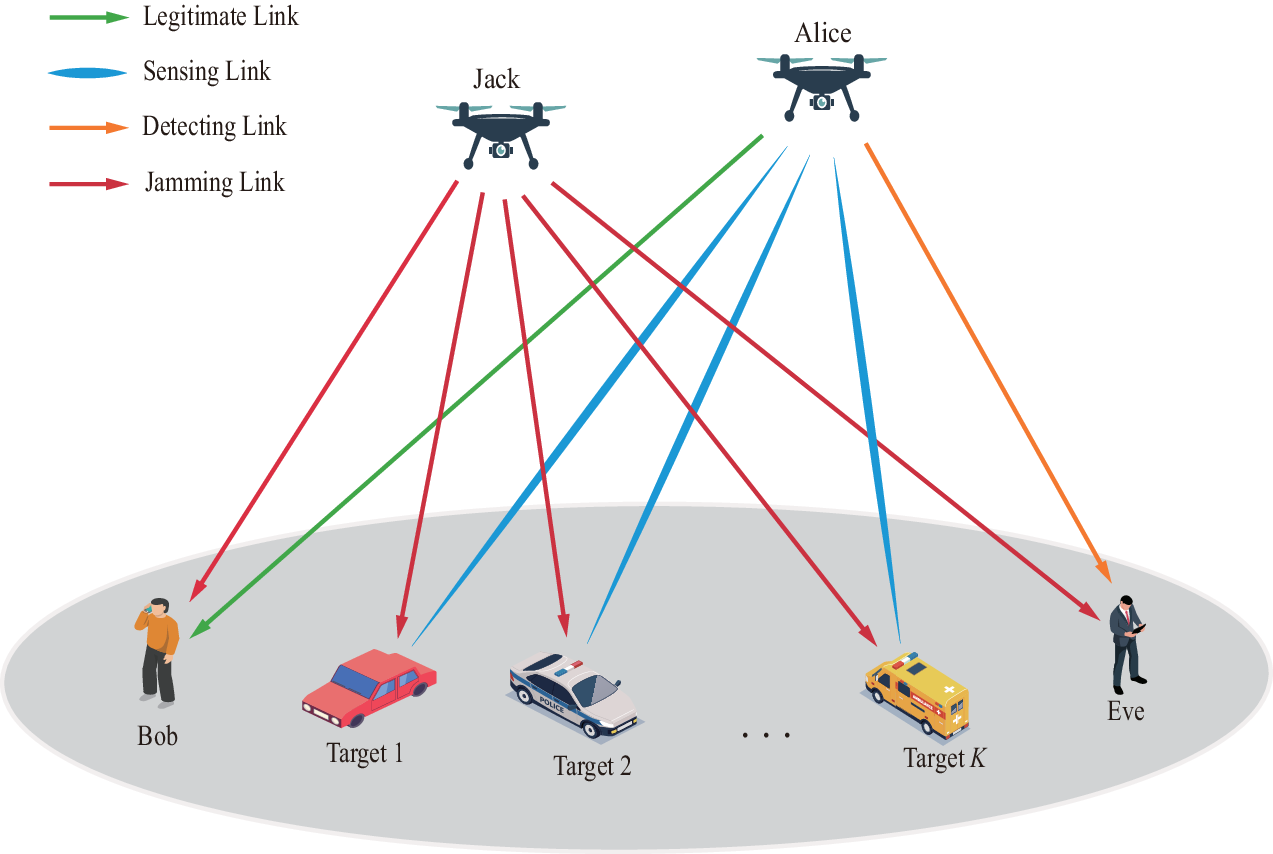}
    \caption{Dual-UAV-enabled A2G-ISAC system model}
    \label{fig:system_model}
\end{center}
\vspace{-0.3in}
\end{figure}

The considered A2G-ISAC system consists of a source UAV (Alice), a jamming UAV (Jack), a legitimate ground user (Bob), an eavesdropper (Eve), and $K$ ground targets, as shown in Fig. \ref{fig:system_model}. The dual-functional aerial BS, i.e., Alice,  transmits confidential signals to Bob while conducting the radar sensing of $K$ ground targets. To enhance both the secure communication and target sensing performance, Jack transmits AN signals to interfere with Eve and aid Alice on detecting the ground targets. Specifically, the target-echoed ISAC waveforms transmitted by Alice and AN signals transmitted by Jack are combined at the aerial BS, i.e., Alice, for detecting the target, as conducted in a hybrid monostatic-bistatic radar system \cite{Hybrid_Bistatic_Monostatic,Hybrid_Multistatic}.  
In the considered system, Alice and Jack are assumed to be respectively equipped with a uniform linear array (ULA) consisting of $M$ antenna elements, while Bob and Eve are respectively equipped with a single receive-antenna.

To realize the secure transmission and target sensing, the dual-UAV operates for a finite task period
$\mathcal{T}\triangleq [0,T]$ ($T> 0$), which is divided into $N$ time slots with a duration $\Delta_t = T/ N$  for each time slot, where $\Delta_t$ is chosen as small as possible to ensure that the dual-UAV location remains constant within a time slot, which will facilitate the joint design of the dual-UAV trajectory and dual-UAV beamforming. In time slot $n \in \mathcal{N} \triangleq \{1, 2, \ldots, N\}$, UAV $m$ maneuvers at the location $\big[{x}_{m}[n],{y}_{m}[n],{H}_m \big]$, where $m\in\{a,j\}$ represents Alice and Jack, respectively, and $H_m$ denotes the minimum safe flight altitude. Moreover, the horizontal location of UAV $m$ is denoted by ${\bm{u}}_{m}[n]= \big[{x}_{m}[n],{y}_{m}[n] \big]^H$. The location of ground node $q$ is represented by ${\bm{v}}_{q}=[{x}_{q},{y}_{q}]^H$, where $q \in \{b, e\}$ represents Bob and Eve, respectively. The initial and final horizontal locations of UAV $m$ are respectively defined as ${\bm u}_m^{_{\rm I}} = [x^{_{\rm I}}, y^{_{\rm I}}]$ and ${\bm u}_m^{_{\rm F}} = [x^{_{\rm F}}, y^{_{\rm F}}]$. Let $V_{\max} = \tilde{V}_{\max} \Delta_t$ specify the maximum UAV displacement in each time slot, where $\tilde{V}_{\max}$ denotes the maximum flight speed of each UAV. During a task period, the locations of the dual-UAV are limited by the following UAV maneuvering constraints
\begin{eqnarray}
  & & ~~~~~~~~ {\bm u}_m[0] = {\bm u}_m^{_{\rm I}},~  {\bm u}_m[N] =  {\bm u}_m^{_{\rm F}},  \label{position_I_F} \\
& & \Vert \bm{u}_m[n]-\bm{u}_m[n-1]\Vert \le V_{\max}~, ~\forall n \in\mathcal{N} . ~~~~ 
\label{displacement}
\end{eqnarray}
Within a task period, the total time slots are classified into two phases, namely, the SC and SCS phases, which are denoted as $\mathcal{N}_c \triangleq \{1,2\cdots,N_c\}$ and $\mathcal{N}_s \triangleq \{1,2\cdots,N_s\}$, respectively, where $N_c$ + $N_s$ = $N$. In the SC phase, Alice and Jack collaborate to enhance the secure communication performance. In the SCS phase, Alice and Jack conduct the secure communication and target sensing cooperatively.  
In time slot $n$, the signals transmitted by Alice and Jack can be respectively expressed as
\begin{equation}
    {\bm{x}}_a[n] = \left\{ {\begin{array}{*{20}{c}}
{ {\bm{w}_{a}}[n]s_{a}[n] },& {\forall n \in {\cal N}_c},\\
{ {\bm{w}_{a}}[n]s_{a}[n]+ \bm{s}_r[n] },&{\forall n \in {\cal N}_s},
\end{array}} \right.
 \label{eq:transmit_sensing_signal}    
\end{equation}
and
\begin{eqnarray}
 {\bm{x}}_{j}[n]={\bm{w}}_{j}[n]{s}_{j}[n] , \qquad   \forall \mathit{n} \in \mathcal{N},
 \label{eq:transmit_AN}
\end{eqnarray}
where ${\bm{w}}_{a}[n]\in {\mathbb{C}}^{M\times1}$ and ${\bm{w}}_{j}[n]\in {\mathbb{C}}^{M\times1}$ are the transmit beamformers at Alice and Jack, respectively, ${s}_{a}[n] \backsim \mathcal{C}\mathcal{N}(0,1)$ is Alice's information signal intended to Bob, ${s}_{j}[n]\backsim \mathcal{C}\mathcal{N}(0,1)$ is the AN signal transmitted by Jack, and ${\bm{s}}_{r}[n]\in {\mathbb{C}}^{M\times1}$ is the sensing signal with zero mean and covariance matrix ${\bm{R}}_{r}[n]=\mathbb{E}( {\bm{s}}_{r}[n]{\bm{s}}_{r}^{H}[n] )\succeq 0$. We assume that ${s}_{a}[n]$ and ${\bm{s}}_{r}[n]$ are independent of each other.  
Based on \eqref{eq:transmit_sensing_signal} and \eqref{eq:transmit_AN}, ${s}_{a}[n]$ and ${s}_{j}[n]$ are respectively transmitted using a single beam,  while ${\bm{s}}_{r}[n]$ is transmitted using ${M}_{s}$ beams, where $0\leq {M}_{s}\leq M$. Specifically, the ${M}_{s}$ beams are generated by performing eigenvalue decomposition on ${\bm{R}}_{r}[n]$ with ${\rm{rank}}(\bm{R}_r[n])=M_s$. The transmit powers of Alice and Jack satisfy 
\begin{equation}
\!\!\!\!\mathbb{E}(\|\bm{x}_a[n]\|^2) \!=\! \left\{ {\begin{array}{*{20}{c}}
\Vert\bm{w}_a[n]\Vert^2\le P_a^{\rm{max}},   n\in\mathcal{N}_c,\\
\Vert\bm{w}_a[n]\Vert^2+ {\rm{tr}}(\bm{R}_r[n])\le P_a^{\rm{max}},   n\in\mathcal{N}_s,
\end{array}} \right.
 \label{power_alice}    
\end{equation}
and
\begin{eqnarray}
\mathbb{E}(\|\bm{x}_j[n]\|^2)=\Vert\bm{w}_j[n]\Vert^2\le P_j^{\rm{max}},~~   n\in\mathcal{N} ,
 \label{power_jack}
\end{eqnarray}
 where $\mathbb{E}(\|\bm{x}_m[n]\|^2)$ and $P_m^{\max}$ denote the average transmit power and the maximum transmit power of UAV $m$ with $m\in\{a,j\}$.

\subsection{Secure Communication Model}

Considering that minimum maneuvering altitude of each UAV is relatively high, 
in general strong LoS links exist between UAVs and ground nodes  \cite{UAV_Throughput,ISAC_UAV_Cooperative_Detection_NETWORK,LoS_3D_trajectory}. Therefore, 
we adopt the LoS channel model in this work and the channel from UAV $m$ ($m\in\{a,j\}$) to ground node $q$ ($q\in\{b,e\}$) is represented as  
\begin{eqnarray}
{\bm{h}}_{mq}[n]&\!\!\!\!\!=\!\!\!\!\!&\bm{a}_{mq}[n]\sqrt{\beta {d}^{-2}_{mq}[n]}\nonumber \\ 
&\!\!\!\!\!=\!\!\!\!\!&\bm{a}_{mq}[n]\sqrt{\frac{\beta }{{\|{\bm{u}}_{m}[n]-{\bm{v}}_{q}\|}^{2}+{H}_{m}^{2}}}, 
\label{communication_channel}
\end{eqnarray}
where $\beta$ denotes the path-loss at the reference distance of ${d}_{0} = 1$ m and ${d}_{mq}[n]=\sqrt{{{\|\bm{u}_{m}[n]-\bm{v}_{q}\|}^{2}+{H}_{m}^{2}}}$ represents the distance between UAV $m$ and ground node $q$. In \eqref{communication_channel},  $\bm{a}_{mq}[n]$ is the steering vector, which is given by
\begin{eqnarray}
\bm{a}_{mq}[n]\!=\!{\!\left[ 1,{e}^{\mathfrak{j}2\pi \frac{{d}_{m}}{\lambda}\cos\theta_{mq}[n]} , \!\cdots \! ,{e}^{\mathfrak{j}2\pi \frac{{d}_{m}}{\lambda}(M-1)\cos\theta_{mq}[n] }\right]\!}^{T}\!\!\!,\!\!\!
\label{steering_vector_mq}
\end{eqnarray}
where $d_m$ and $\lambda$ denote the spacing between adjacent antennas and the wavelength of the carrier signal, respectively, and $\theta_{mq}[n]$ is the angle of departure (AoD) from UAV $m$ to ground node $q$, which can be expressed as
\begin{eqnarray}
    \theta_{mq}[n]= \arccos \frac{{H}_{m}}{\sqrt{{{\|\bm{u}_{m}[n]-\bm{v}_{q}\|}^{2}+{H}_{m}^{2}}}}.
    \label{AoS_lq}
\end{eqnarray} 
In both the SC and SCS phases, the received signal at Bob in time slot $n$  can be written as
\begin{eqnarray}
\!\!\!\!\!\!\!\!y^c_b[n] &\!\!\!\!=& \!\!\! \bm{h}_{ab}[n] \bm{x}_a [n]+\bm{h}_{jb}[n] \bm{x}_j[n]+ {z}_{b}[n] \nonumber \\
&\!\!\!\! =&  \!\!\!\bm{h}_{ab}[n]\bm{w}_a [n]+\bm{h}_{jb}[n]\bm{w}_j[n] s_j[n]+z_b[n]
\label{receive_signal_c}
\end{eqnarray}
and
\begin{eqnarray}
\!\!\!\!\!\!\!\!y^s_b[n] &\!\!\!\!=& \!\!\! \bm{h}_{ab}[n] \bm{x}_a [n]\!+\! \bm{h}_{jb}[n] \bm{x}_j[n]+ {z}_{b}[n] \nonumber \\
&\!\!\!\! =&  \!\!\!\bm{h}_{ab}[n](\bm{w}_a [n]s_a[n]\!+\!\bm{s}_r[n])\!+\!\bm{h}_{jb}[n]\bm{w}_j[n] s_j[n] \nonumber \\
&\!\!\!\!  &  \!\!\!  + z_b[n],
\label{receive_signal_s}
\end{eqnarray}
respectively, where ${z}_{b}\left [ n\right]\backsim \mathcal{C}\mathcal{N}( 0,{\sigma_b }^{2})$ is the additive white Gaussian noise (AWGN) at Bob. 

In the SC phase, the AN signal is first detected and removed from the received signal by using SIC at Bob. Then, the intended information signal ${{s}}_a$ is detected. In the SCS phase, both the AN and sensing signals are first detected before detecting ${{s}}_a$ at Bob. However, residual interference is unavoidable due to imperfect SIC in practice, which results in a degraded system performance \cite{SIC_Rate-Splitting,SIC_NOMA,Beamforming_SIC_ISAC}. To accurately model the residual interference effect, the received SINR at ground node $q$ ($q\in\{b,e\}$) in the SC and SCS phases are respectively expressed as 
\begin{eqnarray}
{\gamma}^c_{q}[n] =   \frac{|\bm{h}_{aq}^H[n]\bm{w}_a[n]|^2}{ \varphi_{jq}|\bm{h}_{jq}^H[n]\bm{w}_j[n]|^2\!+\! \sigma_q^2}
\label{R^c_b}
\end{eqnarray}
and
\begin{eqnarray}    
   \gamma^s_q[n]=\frac{|\bm{h}_{aq}^H[n]\bm{w}_a[n]|^2}{\varphi_{rq}\bm{h}_{aq}^H[n]\bm{R}_r[n]\bm{h}_{aq}[n] + \varphi_{jq} |\bm{h}_{jq}^H[n]\bm{w}_j[n]|^2 + \sigma_q^2}, \!\!\!\!\!\!\!\!\!\!\!\!\nonumber \\
\!\!\!\!\!\!\!\!\!\ 
    \label{SINR_b} 
\end{eqnarray}
where $\varphi_{jq}$ ($0 \leq \varphi_{jq} \leq 1$) and $\varphi_{rq}$ ($0 \leq \varphi_{rq} \leq 1$) denote the residual interference levels after eliminating the AN signal and sensing signal, respectively. Then, for the considered A2G-ISAC system, the secrecy rates in the SC and SCS phases are respectively given by
\begin{eqnarray}
    {R}^c[n]=\big[\log_2(1 + \gamma^c_b[n])-\log_2(1 + \gamma^c_e[n])\big]^+
\label{Ach_rate_n_c}
\end{eqnarray}
and
\begin{eqnarray}
   {R}^s[n]=\big[\log_2(1 + \gamma^s_b[n])-\log_2(1 + \gamma^s_e[n])\big]^+,
\label{Ach_rate_n_s}
\end{eqnarray}
where $[x]^+ \triangleq {\max}(x,0)$ ensures the non-negative secrecy rate.

\subsection{Sensing Performance Metric}

In the SCS phase, Alice and Jack collaborate to sense the $K$ ground targets in addition to conducting secure communication. 
To improve the sense performance, Alice and Jack construct a hybrid monostatic-bistatic radar system, in which the echoed signal arrived at Alice comprises the dual-function waveform transmitted by Alice and the AN signal transmitted by Jack. By combining the high-resolution capabilities of a monostatic radar with the spatial diversity and reduced vulnerability of a bistatic radar, the target detection accuracy and coverage can be enhanced by the proposed scheme \cite{Hybrid_Bistatic_Monostatic,Hybrid_Multistatic}.

The $K$ ground target are randomly located in an area of interest, in which the location of target $k$ is denoted as $\bm{g}_k = [x_k, y_k]^H$, where $k \in \mathcal{K} \triangleq \{1, 2, \ldots, K\}$. Similar to \eqref{communication_channel}, the channel from UAV $m$ to target $k$ can be expressed as
\begin{eqnarray}
\!\!\!\!\!\!\!\!\!{\bm{h}}_{mk}[n]&\!\!\!\!=\!\!\!\!&\bm{a}_{mk}[n] \sqrt{\beta {d}^{-2}_{mk}[n]}\nonumber \\
&\!\!\!\!=\!\!\!\!& \bm{a}_{mk}[n]\sqrt{\frac{\beta }{{\|{\bm{u}}_{m}[n]-{\bm{g}}_{k}\|}^{2}+{H}_{m}^{2}}}, m\in\{a,j\}. 
\label{sensing_channel}
\end{eqnarray}
The incident sensing signal arrived at target $k$ is given by
\begin{eqnarray}
\!\!\!\!\!\!\!\!y_k^{s}[n] &\!\!\!\!\!=&\!\!\!\! \bm{h}_{ak}[n] \bm{x}_a [n]\!+\! \bm{h}_{jk}[n] \bm{x}_j[n] \nonumber \\
&\!\!\!\!\!=& \!\!\!\!\bm{h}_{ak}[n](\bm{w}_a [n]s_a[n]\!+\!\bm{s}_r[n])\!+\!\bm{h}_{jk}[n]\bm{w}_j[n] s_j[n].~
\end{eqnarray}
Then, the incident signal $y_k^{s}[n] $ is reflected back to Alice and Alice employs multi-angle signal fusion to detect target $k$ as performed in a hybrid monostatic-bistatic radar system.  
Unlike conventional monostatic or bistatic radar systems containing a single transmitter, the proposed scheme introduces a cooperative mechanism that takes advantage of both the source and jamming UAVs, so that both the dual-functional waveform and the jamming signal are fully exploited for target sensing. 
To measure the radar sensing performance, the Cram\'{e}r-Rao bound (CRB) is preferred in various ISAC systems \cite{ren2022fundamental}. However, the CRB of the considered hybrid monostatic-bistatic radar system involves the dual-UAV geometry, received sensing SINR, and specific waveforms in a complicated way, making it challenge to jointly design the dual-UAV trajectory and dual-UAV beamforming \cite{Hybrid_Multistatic}.  In this work, we adopt a feasible sensing metric that accommodates the beampattern gains from Alice and Jack, such that the contributions of both UAVs on detecting target $k$ can be quantified. Specifically, the distance-normalized sum-beampattern gain is introduced as 
\begin{eqnarray}  
\zeta(\bm{u}_a[n], \bm{u}_j[n], \bm{g}_k) 
&\!\!\!\!\!=\!\!\!\!\!& \frac{\mathbb{E}\big[|\bm{a}^H_{ak}[n]\bm{x}_a[n]|^2\big]}{d^2_{ak}[n]} \!+\! \frac{\mathbb{E}\big[|\bm{a}^H_{jk}[n]\bm{x}_j[n]|^2\big]}{d^2_{jk}[n]}  \nonumber\\
&\!\!\!\!\! = \!\!\!\!\!& \frac{\bm{a}^H_{ak}[n](\bm{w}_{a}[n]\bm{w}_{a}[n]^H+
\bm{R}_r[n])\bm{a}_{ak}[n]}{ \Vert \bm{u}_a[n] - \bm{g}_k \Vert^2 + H_a^2}\nonumber\\
&\!\!\!\! \!\!\!\!& + \frac{\bm{a}^H_{jk}[n]\bm{w}_{j}[n]\bm{w}_{j}[n]^H \bm{a}_{jk}[n]}{\Vert \bm{u}_j[n] - \bm{g}_k \Vert^2 + H_j^2}.
\label{eq:sensing_SINR}
\end{eqnarray} 
On the right-hand-side of \eqref{eq:sensing_SINR}, the first and second terms  are the distance-normalized beampattern gains resulted by Alice and Jack, respectively, which signify the incident signal powers arrived at target $k$.

\subsection{Problem Formulation}

In this work, we aim at designing the dual-UAV trajectory and dual-UAV beamforming to maximize the ASR, while ensuring the sensing performance requirements. Specifically, the ASR maximization problems are formulated for the SC and SCS purposes, respectively, subject to the corresponding system constraints.  

In the SC phase, the objective is to maximize the ASR $R_{\rm{\rm{asr}}}^c=\frac{1}{N_c}\sum_{n=1}^{N_c} {R}^c[n]$ by jointly optimizing the dual-UAV trajectory $\{ \bm{u}_a[n], \bm{u}_j [n]\}$, secure communication beamformer $\{\bm{w}_{a}[n], \bm{R}_r[n]\}$, and jamming beamformer $\{ \bm{w}_{j}[n] \}$. The ASR maximization problem is formulated as
\begin{subequations}\label{eq:P1_sub}
    \begin{align}
        (\text{P}1): &~\mathop{\max}\limits_{\{\bm{w}_{m}[n]\succeq 0, \bm{u}_m[n]\}}{R}_{\rm{asr}}^c[n]\\
        \text{s.t.}&~\Vert\bm{w}_a[n]\Vert^2\le P_{a}^{\rm max},\!\!\!\!\quad\forall n \in\mathcal{N}_c\label{eq:Pa_max_constraint_c},\\
        &~\Vert\bm{w}_j[n]\Vert^2\le P_{j}^{\rm max}, \!\!\!\! \quad \forall n \in\mathcal{N}_c,\label{eq:Pj_max_constraint_c} \\ 
&~\eqref{position_I_F}~\text{and}~\eqref{displacement}.
    \end{align}
\end{subequations}
In problem (P$1$), constraints \eqref{eq:Pa_max_constraint_c} and \eqref{eq:Pj_max_constraint_c} impose the transmit power limits for Alice and Jack, respectively.

In the SCS phase, the transmitted signal of Alice includes the information signal ${\bm{x}}_a[n]$ and the sensing signal ${\bm{x}}_r[n]$. Consequently, the constraint  $\zeta(\bm{u}_a[n], \bm{u}_j[n], \bm{g}_k) \geq \Gamma$ is enforced to ensure the sensing performance requirements, where $\Gamma$ is a required threshold for the distance-normalized sum-beampattern gain. The ASR maximization problem in the SCS phase can be formulated as
\begin{subequations}\label{eq:P2_sub}
    \begin{align}
        &\!\!\!\!(\text{P}2): ~\mathop{\max}\limits_{\{\bm{w}_{m}[n],\bm{R}_r[n]\succeq 0, \bm{u}_m[n]\}}{R}_{\rm{asr}}^s[n]\\
        &\text{s.t.}~~\zeta(\bm{u}_a[n], \bm{u}_j[n], \bm{g}_k) 
        \ge \Gamma, \quad \forall n \in \mathcal{N}_s, ~\forall k \in \mathcal{K},\label{eq:sensing_constraint_s}\\
        &~~~~~\Vert\bm{w}_a[n]\Vert^2+ {\rm{tr}}(\bm{R}_r[n])\le P_{a}^{\rm max},\quad\forall n \in\mathcal{N}_s\label{eq:Pa_max_constraint_s},\\
        &~~~~~\Vert\bm{w}_j[n]\Vert^2\le P_{j}^{\rm max}, \quad \forall n \in\mathcal{N}_s,\label{eq:Pj_max_constraint_s} \\ 
             &~~~~~\eqref{position_I_F}~\text{and}~\eqref{displacement}.
    \end{align}
\end{subequations}
In \eqref{eq:P2_sub}, the ASR is given by $R_{\rm{\rm{asr}}}^s=\frac{1}{N_s}\sum_{n=1}^{N_s} {R}^s[n]$. Moreover, constraint \eqref{eq:Pa_max_constraint_s} is the transmit power budget for Alice in the SCS phase, which is different from constraint  \eqref{eq:Pa_max_constraint_c} in the SC phase.
Unfortunately, both the objective functions in (P$1$) and (P$2$) are non-concave due to the complicated coupling between the dual-UAV trajectory and dual-UAV beamforming. Also, constraint \eqref{eq:sensing_constraint_s} is non-convex due to the coupling of the dual-UAV trajectory and dual-UAV beamforming. Thus, conventional convex optimizzation methods cannot be applied to obtain the optimal solutions for problems (P$1$) and (P$2$). In what follows, we propose the solution approaches for problems (P$1$) and (P$2$), respectively.  

\section{Dual-UAV Trajectory and Beamforming Optimization for The SC Purpose}

The ASR maximization problem (P$1$) requires joint optimization of the dual-UAV trajectory and dual-UAV beamforming. 
In this section, we tackle problem  (P$1$) by decoupling it into three subproblems, including the dual-UAV beamforming optimization, Alice's 
trajectory optimization, and Jack's trajectory optimization. After deriveing the optimal solutions for the subproblems, we propose a BCD algorithm to obtain the sub-optimal solution for  problem  (P$1$). 
  
\subsection{Optimizing Dual-UAV Beamforming}

With any given dual-UAV trajectory $\{\bm{u}_a[n],\bm{u}_j[n]\}$, problem (P$1$) reduces to the dual-UAV beamforming optimization subproblem, which can be formulated as
\begin{subequations}
\begin{align}
(\text{P}3): &~\mathop{\max}\limits_{\{\bm{w}_{a}[n],\bm{w}_{j}[n]\succeq 0\}}{R}_{\rm{asr}}^c(\bm{w}_{a}[n],\bm{w}_{j}[n]) \\		&~\text{s.t.}~\eqref{eq:Pa_max_constraint_c}~\text{and}~\eqref{eq:Pj_max_constraint_c}.
\end{align}    
\end{subequations}
In problem (P$3$), the solutions for the dual-UAV beamformers $\{\bm{w}_a[n], \bm{w}_j[n]\}$ are independent across different time slots. Consequently, the optimization of the beamformers can be performed independently for each time slot. In time slot $n$, $\forall n \in\mathcal{N}_c$, the subproblem of the dual-UAV beamforming optimization can be expressed as 
\begin{subequations}
\begin{align}
(\text{P}4): &\mathop{\max}\limits_{\bm{w}_{a}[n],\bm{w}_{j}[n]\succeq 0} {R}^c(\bm{w}_{a}[n],\bm{w}_{j}[n]) &\\		
\text{s.t.}
        &~\Vert\bm{w}_a[n]\Vert^2\le P_{a}^{\rm{max}},~ \forall n \in \mathcal{N}_c,\label{eq:Pa_max_constraint2} \\
        &~\Vert\bm{w}_j[n]\Vert^2\le P_{j}^{\max},~ \forall n \in \mathcal{N}_c. \label{eq:Pj_max_constraint2}   
\end{align}    
\end{subequations}
However, due to the coupling of the dual-UAV beamformers $\{\bm{w}_a[n], \bm{w}_j[n]\}$ in the objective function of problem  (P$4$), it is impossible to directly obtain the optimal solution. In what follows, we apply SDR and SCA to tackle problem  (P$4$). 

By introducing $\bm{H}_{mq}[n]\!=\!\bm{h}_{mq}[n]\bm{h}_{mq}^H[n]$ and $\bm{W}_{m}[n]=\bm{w}_{m}[n]\bm{w}_{m}^H[n]$, where $\bm{W}_{m}[n]\succeq 0$, ${\rm{rank}}(\bm{W}_{m}[n])=1$, and $m \in \{a, j\}$, the objective function is rewritten as ${R}^c(\bm{W}_{a}[n],\bm{W}_{j}[n])$ shown in \eqref{eq:hat_R[n]_c} at the bottom of the next page. Then,  problem (P$4$) can be rewritten as 
\setcounter{equation}{23}
\begin{subequations}
    \begin{align}
    (\text{P}5):&~\mathop{\max}\limits_{\bm{W}_{a}[n],\bm{W}_{j}[n]\succeq 0}\!\!\!\!\!\!{{R}}^c(\bm{W}_{a}[n],\bm{W}_{j}[n]) \\ 
    \text{s.t.} 
    &~ {\rm{tr}}(\bm{W}_{a}[n])\le P_{a}^{\max},~ \forall n \in \mathcal{N}_c.\label{eq:Pa_trace_constraint} \\
    &~ {\rm{tr}}(\bm{W}_{j}[n])\le P_{j}^{\max},~ \forall n \in \mathcal{N}_c.  \label{eq:Pj_trace_constraint} \\
    &~ {\rm{rank}}(\bm{W}_{m}[n])\le 1. \label{eq:rank_constraint}
    \end{align}
\end{subequations}
Considering the rank-one constraint  \eqref{eq:rank_constraint} and the non-concave objective function in problem (P$5$), we employ SCA to solve it in what follows. Specifically, in iteration $t_1$, local points $\bm{W}_{a}^{(t_1)}[n]$ and $\bm{W}_{j}^{(t_1)}[n]$ are used to construct 
the first-order Taylor (FoT) expansion of $R^c(\bm{W}_{a}[n],\bm{W}_{j}[n])$, which serves as a lower-bound on the secrecy rate, i.e., ${{R}}^c(\bm{W}_{a}[n],\bm{W}_{j}[n]) \ge \bar {{R}}^{c,(t_1)}(\bm{W}_{a}[n],\bm{W}_{j}[n])$. The form of $\bar R^{c,(t_1)}(\bm{W}_{a}[n],\bm{W}_{j}[n])$ is given by
\begin{eqnarray} 
&{\bar{R}^{c,(t_1)}}\!\!\!\!\!\!&(\bm{W}_{a}[n],\bm{W}_{j}[n])\nonumber\\ 
&\!\!\!\!=\!\!\!\!&\!\! \!\!\!\!\log_2\Big({\rm{tr}}(\bm{H}_{ab}[n]\bm{W}_a[n]) 
   + \varphi_{jb}{\rm{tr}}(\bm{H}_{jb}[n]\bm{W}_j[n]) 
    \nonumber\\ 
&&\!\!\!\!\!\! + \sigma_b^2 \Big)+ \log_2\Big(\varphi_{je}{\rm{tr}}(\bm{H}_{je}[n]\bm{W}_j[n]) 
   + \sigma_e^2 \Big) 
   - a^{(t_1)} \nonumber\\ 
&&\!\!\!\!\!\! - {\rm{tr}}\Big(b^{(t_1)}\bm{H}_{ae}[n]\big(\bm{W}_a[n] 
   - \bm{W}_a^{(t_1)}[n]\big)\Big) \nonumber\\ 
&&\!\!\!\!\!\! - {\rm{tr}}\Big(\big(b^{(t_1)}[n]\varphi_{je}\bm{H}_{je}[n] 
   + c^{(t_1)}[n]\varphi_{jb}\bm{H}_{jb}[n]\big) \nonumber\\ 
&&\!\!\!\! \!\!\times \big(\bm{W}_j[n] 
   - \bm{W}_j^{(t_1)}[n]\big)\Big),
\label{eq:Rc_bound}
\end{eqnarray}
where
\begin{eqnarray} 			
{a}^{(t_1)} \!\!&\!\!\!\!\!=\!\!\!\!\!&\! \!\log_2 \!\Big(\!{\rm{tr}}\big(\bm{H}_{ae}[n]\bm{W}_{a}^{(t_1)}[n]\big) 
                     + \varphi_{je}{\rm{tr}}\big(\bm{H}_{je}[n]\bm{W}_j^{(t_1)}[n]\big) \nonumber\\
  &&\!\! + \sigma_e^2 \Big) 
     + \log_2 \!\Big(\!\varphi_{jb}{\rm{tr}}\big(\bm{H}_{jb}[n]\bm{W}_j^{(t_1)}[n]\big) 
                     + \sigma_b^2 \Big),\!\! 
\label{a^{(t_1)}}
\end{eqnarray}
  \begin{eqnarray} 
		{b}^{(t_1)}\!\!&\!\!\!\!\!\!=\!\!\!\!\!\!&\!\!\frac{\log_2(e)}{{\rm{tr}}(\bm{H}_{ae}[n]\bm{W}_{a}^{(t_1)}[n])
   +\varphi_{je}{\rm{tr}}(\bm{H}_{je}[n]\bm{W}_j^{(t_1)}[n]) +\sigma_e^2},\!\!\nonumber \\
     \label{{b}^{(t_1)}}
		\end{eqnarray}
and
  \begin{eqnarray} 
			{c}^{(t_1)}&\!\!\!\!\!\!=\!\!\!\!\!\!&\frac{\log_2(e)}{\varphi_{jb}{\rm{tr}}(\bm{H}_{jb}[n]\bm{W}_j^{(t_1)}[n])
    +\sigma_b^2}.
     \label{{c}^{(t_1)}}
		\end{eqnarray}
By replacing the objective function ${{R}}^c(\bm{W}_{a}[n],\bm{W}_{j}[n])$ with its lower-bound $\bar{{R}}^{c,(t_1)}(\bm{W}_{a}[n],\bm{W}_{j}[n])$, in iteration $t_1$ of SCA, problem (P$5$) can be approximated as 
\begin{subequations}
\begin{align}
&(\text{P}5.t_1): \max_{\bm{W}_{a}[n],\bm{W}_{j}[n]\succeq 0} 
\bar{R}^{c,(t_1)}(\bm{W}_{a}[n],\bm{W}_{j}[n]) 
\label{eq:P5_n_i_beamforming_opt} \\
&\hspace{5em} \text{s.t. } 
\eqref{eq:Pa_trace_constraint},\, \eqref{eq:Pj_trace_constraint},\, 
\text{and } \eqref{eq:rank_constraint}.
\label{eq:P5_n_i_beamforming_constraints}
\end{align}
\end{subequations}
\begin{figure*}[!b]
\hrulefill
		\normalsize
          \setcounter{equation}{22}
        \begin{eqnarray}
{R}^c(\bm{W}_{a}[n],\bm{W}_{j}[n])=\log_2\bigg(1+\frac{{\rm{tr}}(\bm{H}_{ab}[n]\bm{W}_{a}[n])}{\varphi_{jb}{\rm{tr}}(\bm{H}_{jb}[n]\bm{W}_{j}[n])+\sigma_b^2}\bigg)-
\log_2\bigg
(1+\frac{{\rm{tr}}(\bm{H}_{ae}[n]\bm{W}_{a}[n])}{\varphi_{je}{\rm{tr}}(\bm{H}_{je}[n]\bm{W}_{j}[n])+\sigma_e^2}\bigg)
\label{eq:hat_R[n]_c}	
\end{eqnarray}
\setcounter{equation}{31}
\begin{eqnarray}
\hat{R}^c(\bm{u}_a[n]) \!\!&\!\!\!\!\!\!\!\!=\!\!\!\!\!\!\!\!&\!\! \log_2\bigg(\frac{{\rm{tr}}(\bm{W}_a[n]\bm{A}_{ab}[n])}{d_{ab}^2[n]} 
                    + \frac{\varphi_{jb}{\rm{tr}}(\bm{W}_j[n]\bm{A}_{jb}[n])}{d_{jb}^2[n]}
                    + \frac{\sigma_b^2}{\beta} \bigg) 
             + \log_2\bigg( \frac{\varphi_{je}{\rm{tr}}(\bm{W}_j[n]\bm{A}_{je}[n])}{d_{je}^2[n]} 
                    + \frac{\sigma_e^2}{\beta} \bigg) \nonumber \\
&& - \log_2\bigg( \frac{{\rm{tr}}(\bm{W}_a[n]\bm{A}_{ae}[n])}{d_{ae}^2[n]}
                    + \frac{\varphi_{je}{\rm{tr}}(\bm{W}_j[n]\bm{A}_{je}[n])}{d_{je}^2[n]}
                    + \frac{\sigma_e^2}{\beta} \bigg) 
             - \log_2\bigg( \frac{\varphi_{jb}{\rm{tr}}(\bm{W}_j[n]\bm{A}_{jb}[n])}{d_{jb}^2[n]}
                    + \frac{\sigma_b^2}{\beta} \bigg)
\label{eq:R[n]_trajectory_a}
\end{eqnarray}
\end{figure*}To handle the rank-one constraint \eqref{eq:rank_constraint} in problem (P$5$.$t_1$), we propose to add a penalty term $\frac{1}{\iota _1}\sum_{m\in\{a,j\}}\big(\|\bm{W}_m\|_*+\hat{\bm{W}}_m^{(t_1)}\big)$ into the 
objective function, where 
$\iota_1$ is a control factor and $\hat{\bm{W}}_m^{(t_1)} = \Vert\bm{W}_m\Vert_2 - {\rm{tr}}\big({\bm{p}^{{(t_1)}}_{{{\rm{max}},m}}(\bm{p}^{{(t_1)}}_{{{\rm{max}},m}})^H(\bm{W}_m-\bm{W}_m^{(t_1)})}\big
)$ is the FoT expansion at point $\bm{W}^{(t_1)}_m$. In particular, $\bm{p}^{{(t_1)}}_{{{\rm{max}},m}}$ is the eigenvector corresponding to the largest eigenvalue of $\bm{W}^{(t_1)}_m$ and we adopt $\hat{\bm{W}}_m^{(t_1)}$ as an upper bound on $-\|\bm{W}_m\|_2$. When $\iota_1\rightarrow 0$, we have the fact $\|\bm{W}_m\|_*-\|\bm{W}_m\|_2=0$ to satisfy the rank-one constraint ${\rm{rank}}(\bm{W}_{m}[n]) = 1$. Consequently, the approximated optimization problem is formulated as
\setcounter{equation}{29}
\begin{subequations}
    \begin{align}
(\text{P}6.t_1):~&\mathop{\max}\limits_{\bm{W}_{a}[n],\bm{W}_{j}[n]\succeq 0}{\bar{{R}}^{c,(t_1)}(\bm{W}_{a}[n],\bm{W}_{j}[n])} \nonumber\\
     &\qquad\qquad\qquad-\frac{1}{\iota_1}\sum_{m\in\{a,j\}}\big(\Vert \bm{W}_m \Vert_*+\hat{\bm{W}}_m^{(t_1)}\big). \label{eq:P6_n_i_beamforming_opt}\\
     &\qquad\qquad\text{s.t.}~\eqref{eq:Pa_trace_constraint} \text{ and } \eqref{eq:Pj_trace_constraint}.
    \end{align}
\end{subequations}
Now, problem (P$6$.$t_1$) is convex and we can apply a convex solver to obtain the optimal solution for it.

\subsection{Optimizing Alice's Trajectory}

For any given beamformers $\{\bm{W}_a[n], \bm{W}_j[n]\}$ and Jack's trajectory $\{\bm{u}_j[n]\}$, problem (P$1$) is simplified to optimize Alice's trajectory, with the corresponding subproblem formulated as
\begin{subequations}
    \begin{align}
       \!\!\!\! \!\!(\text{P}7):	~&\!\!\!\!\mathop{\max}\limits_{\{{\bm{u}_a[n]}\}}\frac{1}{N_c}\sum_{n=1}^{N_c}{R}^c(\bm{u}_a[n])
        \label{eq:30a}\\ 
 &\text{s.t.}
~\eqref{position_I_F}~\text{and}~\eqref{displacement}.
    \end{align}
\end{subequations}
However, the complicated coupling between Alice's trajectory and the steering vector in the con-concave objective function prohibits further tackling of problem (P$7$). In this subsection, we first transfer the objective function to a tractable form. Then, we propose a trust-region SCA approach to optimize Alice's trajectory. 

To decouple the Alice's trajectory and steering vector in the objective function of problem (P$7$), the secrecy rate is first rewritten as $\hat{{R}}^c(\bm{u}_a[n]) $ in \eqref{eq:R[n]_trajectory_a} illustrated at the bottom of this page, where  $\bm{A}_{mq}[n]= {\bm{a}}_{mq}[n] {\bm{a}}^H_{mq}[n]$ with $m\in \{a, j\}$ and $q\in \{b, e\}$.  
However, Alice's trajectory $\bm{u}_a[n]$ and steering vector $\bm{a}_{aq}[n]$ are still coupled in $\bm{A}_{aq}[n]$, which makes the secrecy rate a non-concave function. To formulate a tractable objective function, we further rewrite the secrecy rate as
\setcounter{equation}{32}
\begin{align}
\hat{R}^c(\bm{u}_a[n]) &= \log_2(\zeta_{1b}[n]) + \log_2(\zeta_{2e}[n]) \nonumber \\
&\quad - \log_2(\zeta_{1e}[n]) - \log_2(\zeta_{2b}[n]),
\label{eq:R[n]_trajectory_alice}
\end{align}     
where
\begin{eqnarray} 
\zeta_{1q}[n] = \eta_{aq}[n]+\zeta_{2q}[n]
        \label{eq:u_a_zeta}
\end{eqnarray}
and 
\begin{eqnarray}
 \zeta_{2q}[n] =\bigg(\frac{\varphi_{jq}{\rm{tr}}\big(\bm{W}_j[n]\bm{A}_{jq}[n]\big)}{{d}^2_{jq}[n]}+\frac{\sigma_q^2}{\beta}\bigg) d^2_{aq}[n] \label{eq:u_a_tau}   
\end{eqnarray}
with
\begin{eqnarray}
\eta_{mq}[n]\!\!&\!\!\!\!\!=\!\!\!\!\!&\!\!\sum_{y=1}^M\big[\bm{W}_m[n]\big]_{y,y}+2\sum_{x=1}^{M}\sum_{y=x+1}^{M}\big\vert\big[\bm{W}_m[n]\big]_{x,y}\big\vert \nonumber\\
		&&\times \cos\bigg(\theta^{\bm
  {W}_m}_{x,y}[n]+2\pi\frac{d_m}{\lambda}\frac{H_a(y-x)}{d_{mq}[n]}\bigg).
  \label{eq:eta_q_Wa[n]}
	\end{eqnarray}
Then, the secrecy rate is approximated by using the FoT expansion of $ \hat{R}^c(\bm{u}_a[n])$ at local point $\bm{u}_a^{(t_2)}[n]$. Specfically, in iteration $t_2$ of the trust-region SCA, the approximated secrecy rate is given by 
\begin{eqnarray}
		{\hat{R}}^c(\bm{u}_a[n]) 
		&\!\!\!\! \ge \!\!\!\!&  \alpha_a^{(t_2)}[n]+\big({\bm{\rho}}^{(t_2)}_a[n]\big)^H\big(\bm{u}_a[n]-\bm{u}_a^{(t_2)}[n]\big) ~~~~ \nonumber\\
		&\!\!\!\! \triangleq \!\!\!\!& {\tilde{R}}^{c,(t_2)} (\bm{u}_a[n]), \label{eq:u_a_R[n]_ft_ua}
	\end{eqnarray}
where 
  \begin{eqnarray} 	\alpha_a^{(t_2)}\!\!&\!\!=\!\!&\!\!\log_2\big(\zeta_{1b}^{(t_2)}[n]\big)+\log_2\big(\zeta^{(t_2)}_{2e}[n]\big)\nonumber\\
  &&\!\!-\log_2\big(\zeta^{(t_2)}_{1e}[n]\big)-\log_2\big(\zeta^{(t_2)}_{2b}[n]\big)
   \label{eq:R[n]_a_alpha}
\end{eqnarray}
and
\begin{eqnarray} 
\bm{\rho}_a^{(t_2)} \!\!\!&\!\!\!= \!&\!\!\!\! \log_2(e)\Bigg(\!\!\bigg(\frac{\bm{\gamma}_{ab}^{(t_2)}[n]}{\zeta_{1b}^{(t_2)}[n]} \!-\! \frac{\bm{\gamma}_{ae}^{(t_2)}[n]}{\zeta_{1e}^{(t_2)}[n]}\bigg) \!+\! \bigg(\frac{1}{\zeta_{1b}^{(t_2)}[n]} \!-\! \frac{1}{\zeta_{2b}^{(t_2)}[n]} \bigg) \nonumber\\ 
&&\!\!\!\times \Big(\frac{\varphi_{jb}{\rm{tr}}(\bm{W}_j[n]\bm{A}_{jb}[n])}{d^2_{jb}[n]} \!+\! \frac{\sigma_b^2}{\beta}\Big)(\bm{u}_a^{(t_2)}[n] \!-\! \bm{v}_b) \nonumber \\
&&\!\!\!- \bigg(\frac{1}{\zeta_{1e}^{(t_2)}[n]} \!-\! \frac{1}{\zeta_{2e}^{(t_2)}[n]} \bigg)\Big(\frac{\varphi_{je}{\rm{tr}}(\bm{W}_j[n]\bm{A}_{je}[n])}{d^2_{je}[n]} \!+\! \frac{\sigma_e^2}{\beta}\Big) \nonumber \\ 
&&\!\!\!\times (\bm{u}_a^{(t_2)}[n] \!-\! \bm{v}_e)\!\!\Bigg),
\label{eq:R[n]_a_rho}
\end{eqnarray}
with
\begin{eqnarray}
\bm{\gamma}_{aq}^{(t_2)}[n]		\!\!&\!\!\!\!\!\!\!\!\!\!\!\!\!\!\!\! =\!\!\!\!\!\!\!\! \!\!\!\!\!\!\!\!& \!\!\frac{4\pi d_a H_a}{\lambda}\sum_{x=1}^{M}\sum_{y=x+1}^{M}\vert[\bm{W}_a[n]]_{x, y}\vert \sin\bigg(\theta^{\bm{W}_a}_{x,y}[n] \nonumber\\
		&&\! + \frac{2\pi d_a}{\lambda}\frac{H_a (y - x)}{ d_{aq}^{(t_2)}[n]}\bigg)\frac{ (y - x)(\bm{u}^{(t_2)}_a[n]-\bm{v}_q)}{(d_{aq}^{(t_2)}[n])^3}. ~~~~
  \label{eq:u_a_gamma}
	\end{eqnarray}
To ensure the approximation accuracy of the secrecy rate during the iterations of the trust-region SCA approach, the update of Alice's trajectory is constrained by 
\begin{eqnarray}
    \big\Vert \bm{u}_a^{(t_2)}[n]-\bm{u}_a^{({t_2}-1)}[n] \big\Vert \le \psi^{(t_2)}_a , ~~~\forall n\in \mathcal{N}_c. \label{eq:u_a_trust_region}
\end{eqnarray}
In \eqref{eq:u_a_trust_region}, $\psi^{(t_2)}_a$ is the radius of the trust-region in iteration $t_2$, which is updated by $\psi^{({t_2}+1)}_a = r_a{\psi^{(t_2)}_a}$, where $0<r_a<1$ is the factor that controls the convergence speed of the trust-region SCA approach. With the formulated secrecy rate approximation in \eqref{eq:u_a_R[n]_ft_ua} and the introduced trajectory trust-region constraint in \eqref{eq:u_a_trust_region}, the subproblem of optimizing Alice's trajectory can be formulated as 
\begin{subequations}\label{eq:uav_t_a_trajectory_opt}
\begin{align}
		(\text{P8.}t_2): & \mathop{\max}\limits_{\{{\bm{u}_a[n]}\}}  \frac{1}{N_c} \sum_{n=1}^{N_c}{\tilde{R}}^{c,(t_2)}(\bm{u}_a[n])  \\
			&\text{s.t.}~ \eqref{position_I_F},~\eqref{displacement},
   \text{ and } 
   \eqref{eq:u_a_trust_region}.
\end{align}
\end{subequations}  
 Now,  standard convex optimization solvers can be applied to obtain the optimal solution for problem (P$8$.$t_2$). 
 By setting a sufficiently small trust-region radius, the optimal solution for problem (P$8$.$t_2$) can be arrived in convergency.  
 
\subsection{Optimizing Jack's Trajectory}

With the given beamformers $\{\bm{W}_a[n], \bm{W}_j[n]\}$ and Alice's trajectory $\{\bm{u}_a[n]\}$, the ASR maximization problem (P$1$) is reduced to 
\begin{subequations}
\begin{align}
 \!\!\!\!       \!\!(\text{P}9):	~&\!\!\!\!\mathop{\max}\limits_{\{{\bm{u}_j[n]}\}}\frac{1}{N_c}\sum_{n=1}^{N_c}{R}^c(\bm{u}_j[n])
        \label{eq:51a}\\ 
 &\text{s.t.}~
\eqref{position_I_F}~\text{and}~\eqref{displacement}.
\end{align}
\end{subequations}
Obviously, a similar coupling exists between Jack's trajectory and steering vector in the objective function as that in problem (P$7$). Therefore, a similar trust-region SCA approach can be applied to address the problem (P$7$).

Following a similar procedure as that in subsection III.B, the approximated secrecy rate in iteration $t_3$ of the trust-region SCA approach can be written as
\begin{eqnarray}
		\breve{R}^{c,(t_3)} (\bm{u}_j[n])
		&\!\!\!\! \triangleq \!\!\!\!&  \alpha_j^{(t_3)}[n]+\big({\bm{\rho}}^{(t_3)}_j[n]\big)^H\big(\bm{u}_j[n]-\bm{u}_j^{(t_3)}[n]\big),  \nonumber\\
       &\!\!\!\!  \!\!\!\!&  \label{eq:u_a_R[n]_ft_uj}
\end{eqnarray}
where 
 \begin{eqnarray} 	\alpha_j^{(t_3)}\!\!&\!\!=\!\!&\!\!\log_2\big(\zeta_{3b}^{(t_3)}[n]\big)+\log_2\big(\zeta^{(t_3)}_{4e}[n]\big)\nonumber\\
  &&\!\!-\log_2\big(\zeta^{(t_3)}_{3e}[n]\big)-\log_2\big(\zeta^{(t_3)}_{4b}[n]\big)
   \label{eq:R[n]_j_alpha}
\end{eqnarray}
and
\begin{eqnarray}    
\bm{\rho}_j^{(t_3)}\!\!\!&\!\!\!=\!\!\!&\!\!\log_2(e)\Bigg(\!\!\bigg (\frac{1}{\zeta 
 _{3b}^{(t_3)}[n]}\!-\!\frac{ 1}{\zeta  _{4b}^{(t_3)}[n]} \bigg)\!\bigg(\varphi_{jb}\bm{\gamma}_{jb}^{(t_3)}[n]+\frac{\sigma_b^2}{\beta}\nonumber\\
 \!\!\!\!&&\!\!\times(\bm{u}_j^{(t_3)}[n]-\bm{v}_b)\!\!\bigg)\!+\!\frac{{\rm{tr}}\big(\bm{W}_a[n]\bm{A}_{ab}[n]\big)(\bm{u}_j^{(t_3)}[n]-\bm{v}_b)}{{d}^2_{ab}[n]\zeta 
 _{3b}^{(t_3)}[n]}\nonumber\\
 \!\!\!\!&&\!\!-\bigg (\frac{1}{\zeta 
 _{3e}^{(t_3)}[n]}\!-\!\frac{ 1}{\zeta  _{4e}^{(t_3)}[n]} \bigg)\!\bigg(\varphi_{je}\bm{\gamma}_{je}^{(t_3)}[n]+\frac{\sigma_e^2}{\beta}\nonumber\\
\!\!\!\!&&\!\!\times(\bm{u}_j^{(t_3)}[n]-\bm{v}_e)\!\!\bigg)\nonumber\\
 \!\!\!\!&&\!\!-\frac{{\rm{tr}}\big(\bm{W}_a[n]\bm{A}_{ae}[n]\big)(\bm{u}_j^{(t_3)}[n]\!-\!\bm{v}_e)}{{d}^2_{ae}[n]\zeta 
 _{3e}^{(t_3)}[n]}\Bigg), 
 \label{eq:R[n]_j_rho}
		\end{eqnarray}
with
\begin{eqnarray}
\bm{\gamma}_{jq}^{(t_3)}[n]
		\!\!&\!\!\!\!\!\!\!\!\!\!\!\!\!\!\!\! =\!\!\!\!\!\!\!\! \!\!\!\!\!\!\!\!& \!\!\frac{4\pi d_j H_j}{\lambda}\sum_{x=1}^{M}\sum_{y=x+1}^{M}\big\vert[\bm{W}_j[n]]_{x, y}\big\vert \sin\bigg(\theta^{\bm{W}_j}_{x,y}[n] \nonumber\\
		&&\! + \frac{2\pi d_j}{\lambda}\frac{H_j (y - x)}{ d_{jq}^{(t_3)}[n]}\bigg)\frac{ (y - x)(\bm{u}^{(t_3)}_j[n]-\bm{v}_q)}{(d_{jq}^{(t_3)}[n])^3}. ~~~~
  \label{eq:u_j_gamma}
	\end{eqnarray}
Furthermore, the update of Jack's trajectory in iteration $t_3$ is constrained by 
\begin{eqnarray}
    \big\Vert \bm{u}_j^{(t_3)}[n]-\bm{u}_j^{(t_3-1)}[n] \big\Vert \le \psi^{(t_3)}_j , ~~~\forall n\in \mathcal{N}_c. \label{eq:u_j_trust_region}
\end{eqnarray}
In \eqref{eq:u_j_trust_region}, the trust-region radius is updated by   $\psi^{({t_3}+1)}_j = r_j{\psi^{(t_3)}_a}$, where $0<r_j<1$ is the factor to control the convergence speed. Then, the subproblem of optimizing Jack's trajectory can be reformulated as  
\begin{subequations}\label{eq:uav_t_j_trajectory_opt}
		\begin{align}
			\!\!\!\!(\text{P10.}t_3): & \mathop{\max}\limits_{\{{\bm{u}_j[n]}\}} 
   \frac{1}{N_c}\sum_{n=1}^{N_c} {\breve{R}}^{c,(t_3)}(\bm{u}_j[n])  \\
			&\text{s.t.}~ \eqref{position_I_F},~\eqref{displacement},
			\text{ and } \eqref{eq:u_j_trust_region}.
		\end{align}
	\end{subequations}
Now, problem (P$10$.$t_3$) can be effectively addressed by using standard convex optimization solvers.
 
\subsection{BCD Algorithm }

To obtain a suboptimal solution for problem (P$1$), a BCD algorithm is proposed as summarized in Algorithm 1. In the proposed BCD algorithm,  the dual-UAV trajectory and beamformers are iteratively updated. Specifically, problem (P$6$.$t_1$) is solved to optimize the  beamformers $\bigl\{\bm{W}^{(t_1)}_m[n]\bigr\}$ with any given dual-UAV trajectory $\bigl\{\bm{u}^{(t_0)}_a[n], \bm{u}^{(t_0)}_j[n]\bigr\}$. Then, 
problem (P$8$.$t_2$) is solved to optimize  Alice's trajectory $\bigl\{\bm{u}^{(t_2)}_a[n]\bigr\}$ with any fixed $\bigl\{\bm{W}^{(t_0)}_m[n]\bigr\}$ and $\bigl\{\bm{u}^{(t_0)}_j[n]\bigr\}$. As such, problem (P$10$.$t_3$) is solved to optimize Jack's trajectory $\bigl\{\bm{u}^{(t_3)}_j[n]\bigr\}$ with any  given  $\bigl\{\bm{W}^{(t_0)}_m[n]\bigr\}$ and $\bigl\{\bm{u}^{(t_0)}_a[n]\bigr\}$. The iteration of the BCD algorithm terminates when the improvement in the ASR falls below a predefined threshold. 

The computational complexity of the proposed BCD algorithm is analyzed systematically in what follows. For problem (P$6$.$t_1$), which involves $2M^2$ optimization variables and $2$ convex constraints, the computational complexity is $\mathcal{O}\big(\!\log(1/\varrho)N_c(2M^2+2)^{3.5}\big)$, where  $\varrho$ represents the convergence precision. Since problems (P$8$.$t_2$) and (P$10$.$t_3$) have identical structures with $2N_c$ optimization variables and $ N+Nc+1 $ convex constraints, the computational complexity is  $\mathcal{O}\big(\!\log(1/\varrho)(N+3Nc+1)^{3.5}\big)$.

\begin{algorithm}[tt]
    { \caption{BCD Optimization Algorithm for Problem (P$1$) }
        \begin{algorithmic}[1]
            \State \textbf{Initialize}\big\{${\bm{u}}_a^{(t_0)}[n]$\big\}, \big\{${\bm{u}}_j^{(t_0)}[n]$ \big\}, and set $t_0 = 0$.
            \State\textbf{repeat} $t_0\leftarrow t_0+1$
               \State\indent\textbf{repeat}  $t_1\leftarrow t_1+1$ 
            \State\indent\quad\quad  Given  $\bm{u}^{(t_0)}_a[n]$ and $\bm{u}^{(t_0)}_j[n]$, solve problem \Statex\indent\indent\quad\quad (P$6$.$t_1$) to obtain $\bm{W}_m^{(t_1)}[n]$
            \State\indent \textbf{Until}  ${\bar{R}}^{c,(t_1+1)}-{\bar{R}}^{c,(t_1)}\le\varphi_1$
       \State\indent Update $\{{\bm{W}}_m^{(t_0)}[n]\}\!=\!\{{\bm{W}}_m^{(t_1)}[n]\}$ 
       \State\indent \textbf{repeat} $t_2\leftarrow t_2+1$
            \State\indent\quad\quad  Given $\bm{u}^{(t_0)}_j[n]$ and $\bm{W}_m^{(t_0)}[n]$, solve problem
            \Statex\indent\indent\quad\quad  (P$8$.$t_2$) to obtain $\bm{u}_a^{(t_2)}[n]$
            \State\indent\textbf{Until} $\frac{1}{N_c} \sum_{n=1}^{N_c}({\tilde{R}}^{c,(t_2+1)}[n]-{\tilde{R}}^{c,(t_2)}[n])\le\varphi_2$
        \State\indent Update $\bm{u}_{a}^{(t_0)}(n)=\bm{u}_{a}^{(t_2)}(n)$
                \State\indent\textbf{repeat} $t_3\leftarrow t_3+1$
            \State\indent\quad\quad  Given $\bm{u}^{(t_0)}_a[n]$ and $\bm{W}_m^{(t_0)}[n]$, solve problem  
            \Statex\indent\indent\quad\quad (P$10$.$t_3$) to obtain $\bm{u}_j^{(t_3)}[n]$
                   \State \indent\textbf{Until} $\frac{1}{N_c} \sum_{n=1}^{N_c}({\breve{R}}^{c,(t_3+1)}[n]-{\breve{R}}^{c,(t_3)}[n])\le\varphi_3$
        \State\indent Update $\bm{u}_{j}^{(t_0)}(n)=\bm{u}_{j}^{(t_3)}(n)$
            \State\textbf{Until} $R^{c,(t_0+1)}_{\rm{asr}}-R^{c,(t_0)}_{\rm{asr}}\le \varphi_0$
            \State\textbf{Output} final results $\bm{W}_{m}^{(t_0)}(n), \bm{u}_{m}^{(t_0)}(n), \text{and}~\bm{u}_{j}^{(t_0)}(n)$, 
            \Statex $n\in \mathcal{N}_c$.        
        \end{algorithmic}
    }
\end{algorithm}

\section{Dual-UAV Trajectory and Beamforming Optimization for The SCS Purpose}

To accomplish target sensing within a task period, the dual-UAV trajectory and dual-UAV beamforming must be optimized for the SCS purpose. However, determining the optimal dual-UAV trajectory and dual-UAV beamforming by solving problem (P$2$) is not straightforward due to the complicated coupling and non-convex expressions. In this work, we assume that $\Delta_{N_s} = N_s/K$ time slots are allocated to detect each target in the SCS phase. Based on the dual-UAV trajectory optimized in the SC phase, the suitable dual-UAV locations for the SCS purpose can be selected by assuming $N_c = N$, while the dual-UAV beamforming optimization in problem (P$2$) can be separately solved with any given dual-UAV sensing locations. In this section, we first determine the dual-UAV locations suitable for the SCS purpose by formulating  a weighted distance minimization problem and propose a greedy algorithm to solve it. Then, the dual-UAV beamforming optimization is tackled for the SCS purpose with the determined dual-UAV sensing locations.

\begin{algorithm}[t]
{
\caption{Greedy Algorithm for Determining Dual-UAV Sensing Locations}
\begin{algorithmic}[1]
    \State Calculate $\mathcal{D}_k = \{d_k^{\rm{scs}}[1], d_k^{\rm{scs}}[2], \cdots, d_k^{\rm{scs}}[N]\}$, $\forall k \in \{1, \cdots, K\}$; Set $\Delta_{N_s}=N_s/K$ and $k = 1$
    \State \textbf{repeat} $k \leftarrow k + 1$
    \State \indent Sort $\mathcal{D}_k$ by ascending distances  
    \State \indent Select the $\Delta_{N_s}$ time slot indices corresponding
    \Statex \indent\indent to the $\Delta_{N_s}$ smallest distances
    \State \indent Use the selected $\Delta_{N_s}$ time slot indices  
    \Statex \indent\indent to construct the set $\widetilde{\mathcal{D}}_k$    
    \State \indent Exclude $\widetilde{\mathcal{D}}_k$ from $\mathcal{D}_{k+1}, \mathcal{D}_{k+2}, \cdots, \mathcal{D}_K$
    \State \textbf{Until} $k = K$
    \State Based on the time slot indices in $\widetilde{\mathcal{D}}_1, \widetilde{\mathcal{D}}_2, \cdots, \widetilde{\mathcal{D}}_K$, 
    choose the dual-UAV sensing locations from the dual-UAV trajectory optimized for the SC phase. 
\end{algorithmic}
}
\end{algorithm}

\subsection{Optimizing Dual-UAV Sensing Locations}

In the SCS phase, both the secure communication and target sensing performance need to be guaranteed. Specifically, the ASR needs to be maximized, while the distance-normalized sum-beampattern gain must be ensured to be no less than the predefined threshold. Since both the ASR and distance-normalized sum-beampattern gain are closely related to the distances from dual-UAV to ground nodes and targets, a trade-off between the secure communication performance and the sensing performance can be tuned by properly setting the distances of the A2G links. In particular, the distance between Alice and Bob is positive to improve the ASR. On the other hand, a larger distance-normalized sum-beampattern gain corresponds to a shorter distance from either Alice or Jack to target $k$, which is pivotal to ensure the sensing performance. In what follows, a weighted distance minimization problem is formulated to optimize the dual-UAV sensing locations for the SCS purpose.  

To balance the trade-off between the secure communication performance and sensing performance, a weighted distance is introduced as 
\setcounter{equation}{49}
\begin{eqnarray}
d_k^{\rm{scs}}[n]=\tau\big(d_{ak}[n]+d_{jk}[n]\big)+(1-\tau)d_{ab}[n]
, ~~\label{eq:weighted distances}
\end{eqnarray}
where $\tau$ is a weighting factor satisfying $0 \le \tau \le 1$. In \eqref{eq:weighted distances}, the distances $d_{ak}[n]$, $d_{jk}[n]$, and $d_{ab}[n]$ are calculated using the dual-UAV trajectory optimized for the SC purpose, as performed in Section III. Under the assumption of $N_c = N$, equivalently, $\forall n \in {\cal N}_c$, any dual-UAV locations optimized for the SC purpose can be potential candidates suitable for sensing. Equivalently, determining the dual-UAV locations suitable for sensing is transferred to choosing the proper time slots for the dual-UAV to conduct sensing. The corresponding weighted distance minimization problem can be formulated as
\begin{subequations}\label{eq:min weighted distance}
\begin{align}
    (\text{P11}):~ & \mathop{\min}\limits_{\mathcal{N}_s}  \sum_{n\in\mathcal{N}_s}\sum_{k=1}^{K}  d_k^{\rm{scs}}[n]  \\
    & \text{s.t.}~~ |\mathcal{N}_s|=N_s. \label{eq: constrait min weighted distance}
\end{align}
\end{subequations}
\begin{figure*}[!b]
\hrulefill
\normalsize
\setcounter{equation}{52} 
 \begin{eqnarray}
    R^s(\bm{W}_{a}[n],\bm{W}_{j}[n],\bm{R}_{r}[n])\!&\!\!\!\!=\!\!\!\!&\!\log_2\!\bigg(1+\frac{{\rm{tr}}(\bm{H}_{ab}[n]\bm{W}_{a}[n])}{\varphi_{rb}{\rm{tr}}(\bm{H}_{ab}[n]\bm{R}_{r}[n])+\varphi_{jb}{\rm{tr}}(\bm{H}_{jb}[n]\bm{W}_{j}[n])+\sigma_b^2}\bigg)\nonumber\\
   &&-
\log_2\!\bigg
(1+\frac{{\rm{tr}}(\bm{H}_{ae}[n]\bm{W}_{a}[n])}{\varphi_{re}{\rm{tr}}(\bm{H}_{ae}[n]\bm{R}_{r}[n])+{\varphi_{je}\rm{tr}}(\bm{H}_{je}[n]\bm{W}_{j}[n])+\sigma_e^2}\bigg) \label{eq:hat_R[n]}	
\end{eqnarray} 
 
\begin{eqnarray}
{\bar{R}^{s,(t_4)}}(\bm{W}_{a}[n],\bm{W}_{j}[n],\bm{R}_r[n])&\!\!\!\!\!\!=\!\!\!\!\!\!&\log_2\Big({\rm{tr}}(\bm{H}_{ab}[n]\bm{W}_{a}[n])+\varphi_{jb}{\rm{tr}}(\bm{H}_{jb}[n]\bm{W}_{j}[n])+\varphi_{rb}{\rm{tr}}(\bm{H}_{ab}[n]\bm{R}_r[n]) +\sigma_b^2 \Big)\nonumber\\
&&+
\log_2\Big(\varphi_{re}{\rm{tr}}(\bm{H}_{ae}[n]\bm{R}_{r}[n])\nonumber + \varphi_{je}{\rm{tr}}(\bm{H}_{je}[n]\bm{W}_{j}[n]) +\sigma_e^2 \Big)-{a}^{(t_4)}\nonumber\\  
&&-
  {\rm{tr}}\Big({b}^{(t_4)}[n]\bm{H}_{ae}[n]\big(\bm{W}_{a}[n]-\bm{W}_{a}^{(t_4)}[n]\big)\!\Big)\nonumber\\
  &&-{\rm{tr}}\Big(\!\big({b}^{(t_4)}[n]\varphi_{re}\bm{H}_{ae}[n]+{c}^{(t_4)}\varphi_{rb}\bm{H}_{ab}[n]\big)\!\big(\bm{R}_r[n]-\bm{R}_r^{(t_4)}[n]\big)\!\Big)\nonumber\\
  &&-{\rm{tr}}\Big(\!\big({b}^{(t_4)}[n]\varphi_{je}
\bm{H}_{je}[n]+{c}^{(t_4)}[n]\varphi_{jb}\bm{H}_{jb}[n]\big)\big(\bm{W}_j[n]-\bm{W}_j^{(t_4)}[n]\big)\!\Big) 
\label{R[n]_t_4_s}
\end{eqnarray}
\vspace{-0.1in}
		\begin{eqnarray} 
{a}^{(t_4)}&\!\!\!\!=\!\!\!\!&\log_2 \!\Big({\rm{tr}}\big(\bm{H}_{ae}[n]\bm{W}_{a}^{(t_4)}[n]\big) \!+\!
   \varphi_{re}{\rm{tr}}\big(\bm{H}_{ae}[n]\bm{R}_r^{(t_4)}[n]\big)+\varphi_{je}{\rm{tr}}\big(\bm{H}_{je}[n]\bm{W}_j^{(t_4)}[n]\big)\!+\!\sigma_e^2 \Big)\nonumber\\
   &&+\log_2 \!\Big(\varphi_{rb}{\rm{tr}}\big(\bm
{H}_{ab}[n]\bm{R}_{r}^{(t_4)}[n]\big)+
    \varphi_{jb}{\rm{tr}}\big(\bm{H}_{jb}[n]\bm{W}_j^{(t_4)}[n]\big)\!+\!\sigma_b^2 \Big)
\label{{a}^{(t_4)}}		
  \end{eqnarray}
   \vspace{-0.1in}
		\begin{eqnarray}  		
{b}^{(t_4)}=\frac{\log_2(e)}{{\rm{tr}}(\bm{H}_{ae}[n]\bm{W}_{a}^{(t_4)}[n])+\varphi_{re}{\rm{tr}}(\bm{H}_{ae}[n]\bm{R}_r^{(t_4)}[n])
   +\varphi_{je}{\rm{tr}}(\bm{H}_{je}[n]\bm{W}_j^{(t_4)}[n]) +\sigma_e^2}
     \label{{b}^{(t_4)}}
		\end{eqnarray}
   \vspace{-0.1in}
   \begin{eqnarray} 
{c}^{(t_4)}=\frac{\log_2(e)}{\varphi_{rb}{\rm{tr}}(\bm{H}_{ab}[n]\bm{R}_{r}^{(t_4)}[n])+\varphi_{jb}{\rm{tr}}(\bm{H}_{jb}[n]\bm{W}_j^{(t_4)}[n])
    +\sigma_b^2}
     \label{{c}^{(t_4)}}
		\end{eqnarray}
  \end{figure*}Problem (P$11$) indeed is a minimum set cover problem, which is NP-hard to solve. Therefore, we propose a greedy algorithm that minimizes the sum weighted distance $\sum \nolimits_{n\in \mathcal{N}_s} \sum\nolimits_{k = 1}^K d_k^{\rm{scs}}[n]$ to determine the dual-UAV locations suitable for sensing, as summarized in Algorithm 2. In the proposed greedy algorithm, the distances corresponding to target $k$ in all the time slots are calculated and stored in the set  $\mathcal{D}_k =  \{d_k^{\rm{scs}}[1],d_k^{\rm{scs}}[2],\cdots,d_k^{\rm{scs}}[N]\}$. For target 1, the $\Delta_{N_s}$ time slots in $\mathcal{D}_1$ that result in the smallest $\sum\nolimits_{n \in {\cal N}_s} d_1^{\rm{scs}}[n]$ are selected for sensing. Then, the distances of the selected time slots are excluded from the sets $\mathcal{D}_2, \cdots, \mathcal{D}_K$. For target $k$ ($2 \le k \le K$), the $\Delta_{N_s}$ time slots in $\mathcal{D}_k$ that result in the smallest $\sum\nolimits_{n \in {\cal N}_s} d_k^{\rm{scs}}[n]$ are selected for sensing. Then, the distances of the selected time slots are excluded from the sets $\mathcal{D}_{k+1}, \cdots, \mathcal{D}_{N}$. Finally, by using the selected time slots, the dual-UAV sensing locations in the SCS phase are extracted from the dual-UAV trajectory optimized for the SC purpose. 
   
\subsection{Optimizing Dual-UAV Beamforming}

After determining the dual-UAV sensing locations, problem (P$2$) reduces to optimize the dual-UAV beamforming for the SCS purpose. The corresponding ASR maximization subproblem can be expressed as
\setcounter{equation}{51} 
\begin{subequations}
    \begin{align}
    \!\!(\text{P}12):& \!\!\mathop{\max}\limits_{\bm{W}_{a}[n],\bm{W}_{j}[n],\bm{R}_r[n]\succeq 0}\!\!\!\!\!\!{{R}}^s(\bm{W}_{a}[n],\bm{W}_{j}[n],\bm{R}_r[n]) \\ 
    \text{s.t.}~~~~ & \!\!\!\!\!\! \frac{{\rm{tr}}(\bm{A}_{ak}[n]\bm{W}_{a}[n])+{\rm{tr}}(\bm{A}_{ak}[n]\bm{R}_{r}[n])}{d_{ak}^2[n]}
    \nonumber\\
    &\!\!\!\!\!\! + \frac{{\rm{tr}}(\bm{A}_{jk}[n]\bm{W}_{j}[n])}{d_{jk}^2[n]}
    \ge \Gamma, ~\forall n \in \mathcal{N}_s,~\forall k \in \mathcal{K}, \label{eq:sensing_SINR_constraint2_s}\\
    &\!\!\!\!\!\! {\rm{tr}}(\bm{W}_{a}[n])+{\rm{tr}}(\bm{R}_r[n])\le P_{a}^{\max},~\forall n \in \mathcal{N}_s, \label{eq:Pa_trace_constraint_s} \\
    &\!\!\!\!\!\! {\rm{tr}}(\bm{W}_{j}[n])\le P_{j}^{\max},~\forall n \in \mathcal{N}_s, \label{eq:Pj_trace_constraint_s} \\
    &\!\!\!\!\!\! {\rm{rank}}(\bm{W}_{m}[n])\le 1,~\forall n \in \mathcal{N}_s,~m \in \{a, j\}. \label{eq:rank_constraint_s}
    \end{align}
\end{subequations}
For problem (P$12$), we express the secrecy rate ${R}^s(\bm{W}_{a}[n],\bm{W}_{j}[n],\bm{R}_{r}[n])$ as that in \eqref{eq:hat_R[n]}, which is shown at the bottom of this page.
To tackle the non-concave objective function, we apply SCA to obtain a suboptimal solution for problem (P$12$). In particular, the FoT expansion of the secrecy rate is adopted as a lower-bound on the objective function, which is shown in \eqref{R[n]_t_4_s} with $a^{(t_4)}$, $b^{(t_4)}$, and $c^{(t_4)}$ given by \eqref{{a}^{(t_4)}}, \eqref{{b}^{(t_4)}}, and \eqref{{c}^{(t_4)}}, respectively, as illustrated at the bottom of this page. Furthermore, to address the rank-one constraint \eqref{eq:rank_constraint_s}, a penalty term $\frac{1}{\iota _2}\sum_{m\in\{a,j\}}(\|\bm{W}_m\|_*+\hat{\bm{W}}_m^{(t_4)})$ is introduced into the objective function, where 
$\iota_2$ is a penalty parameter, and $\hat{\bm{W}}_m^{(t_4)}$ is
an upper bound on $-\|\bm{W}_m\|_2$, i.e., $-\|\bm{W}_m\|_2 \leq \hat{\bm{W}}_m^{(t_4)}$. Using the FoT expansion at point $\bm{W}^{(t_4)}_m$,  $\hat{\bm{W}}_m^{(t_4)}$ is given by $\hat{\bm{W}}_m^{(t_4)} = \Vert\bm{W}_m\Vert_2 - {\rm{tr}}\big({\bm{o}^{{(t_4)}}_{{{\rm{max}},m}}(\bm{o}^{{(t_4)}}_{{{\rm{max}},m}})^H(\bm{W}_m-\bm{W}_m^{(t_4)})}\big
)$, where $\bm{o}^{(t_4)}_{{{\rm{max}},m}}$ is the eigenvector corresponding to the largest eigenvalue of $\bm{W}^{(t_4)}_m$. 
Then, using SCA, problem (P$12$) can be approximated as  
\setcounter{equation}{57} 
\begin{subequations}
    \begin{align}
(\text{P}13.t_4):~&\!\!\!\!\!\!\!\!\!\mathop{\max}\limits_{\bm{W}_{a}[n],\bm{W}_{j}[n],\bm{R}_r[n]\succeq 0}\!\!\!\!\!\!{\bar{{R}}^{s,(t_4)}(\bm{W}_{a}[n],\bm{W}_{j}[n],\bm{R}_r[n])} \nonumber\\
     &  \qquad\qquad\qquad -\frac{1}{\iota _2}\!\sum_{m\in\{a,j\}}\!\big(\|\bm{W}_m\|_*+\hat{\bm{W}}_m^{(t_4)}\big) \label{eq:P13_n_i_beamforming_opt_s}\\&\text{s.t.}~\eqref{eq:sensing_SINR_constraint2_s},~ \eqref{eq:Pa_trace_constraint_s},\text{ and } \eqref{eq:Pj_trace_constraint_s}.
    \end{align}
\end{subequations}
Now, standard convex optimization tools can be applied to achieve the optimal solution for problem (P$13$.$t_4$). 
 
\section{Simulation Results}

In this section, we present simulation results to evaluate the system performance of the proposed SCS scheme. We consider that $K = 4$ targets are randomly located on the ground. Other simulation parameters are given in Table \ref{tab:simulation_parameters}, unless otherwise specified. For the comparison purpose, the following benchmark schemes are considered in the simulation.  
\begin{itemize}
    \item {Fly-hover-fly (FHF): In the FHF scheme, Alice (Jack) flies from its initial location towards Bob (Eve) at the maximum speed; Then, Alice (Jack) hovers above Bob (Eve) as long as possible;  Alice (Jack) flies towards its final location at the maximum speed within the remained task period. The maximum ratio transmission (MRT) is adopted to realize dual-UAV beamforming with the main-labes of the beams of Alice and Jack toward Bob and Eve, respectively. The dual-UAV sensing locations are determined by using the greedy algorithm.}
    \item {FHF plus beamforming: In this scheme (denoted as ``FHF+Beamforming'' in the remained figures), the dual-UAV follows the trajectory of the FHF scheme, while the dual-UAV sensing locations are obtained by using the greedy algorithm. The dual-UAV beamforming in the SC and SCS phases are obtained by solving problems (P$6$.$t_1$) and (P$13$.$t_4$), respectively.}
    \item {Single-UAV: In the single-UAV scheme, only the source UAV Alice is deployed following the FHF trajectory, while the jamming UAV Jack is omitted. The transmit beamforming of Alice is obtained by using MRT with the main-lobe of the beam toward Bob.}
\end{itemize}

\begin{table}[tt]  
 \scriptsize  
  \renewcommand{\arraystretch}{1.4} 
  \setlength{\tabcolsep}{3pt} 
  \caption{Simulation Parameters}  
  \vspace{-0.2cm}
  \label{tab:simulation_parameters}  
  \begin{tabular}{c|c}  
    \hline\hline
    \bf{Parameter} & \bf{Value} \\  
    \hline 
    Initial location of each UAV & ${\bm u}_a^{_{\rm I}} = {\bm u}_j^{_{\rm I}} = [0,0] $ m \\ \hline  
    Final location of each UAV & ${\bm u}_a^{_{\rm F}}  = {\bm u}_j^{_{\rm F}} = [100,0] $ m \\ \hline 
    Height of each UAV  & $H_a = 120~\text{m}$, ${H}_j  = 100~\text{m}$ \\ \hline 
    Location of Bob & ${\bm v}_b = [40,60]$ m \\ \hline 
    Location of Eve & ${\bm v}_e = [60,60]$ m \\ \hline 
    Maximum speed of each UAV & $V_{\max} = 20~\text{m/s}$ \\ \hline 
    Noise power & $\sigma^2_{b} = \sigma^2_{e} = -80~\text{dBm}$ \\ \hline 
    Path-loss at the reference distance & $\beta^2 = -30~\text{dB}$ \\ \hline 
    Duration of each time slot & $\Delta_t = 0.5$ s \\  
     \hline 
    Number of time slots in the SCS phase & $N_s = 8$ \\
    \hline
    Number of time slots for sensing a target & $\Delta_{N_s} = 2$\\ \hline
    The number of the transmit antennas & $M = 4$ \\ \hline
    Sum-beampattern gain threshold & $\Gamma=-20$ dBm\\ \hline
     Task period duration & $T=10$ s \\ \hline
    Maximum transmit power of each UAV & $P^{\rm{max}}_a=30~\text{dBm}$, $P^{\rm{max}}_j=25~\text{dBm}$ \\ \hline
     Residual interference level at Bob & $\varphi_{rb}=\varphi_{jb} = \varphi = -20$ dB\\
    \hline
    Residual interference level at Eve & $\varphi_{re}=\varphi_{je} = 0$ dB \\
    \hline
  \end{tabular}  
\end{table}

\begin{figure}[tt]
 \begin{center}
    \includegraphics[width=3.2in]{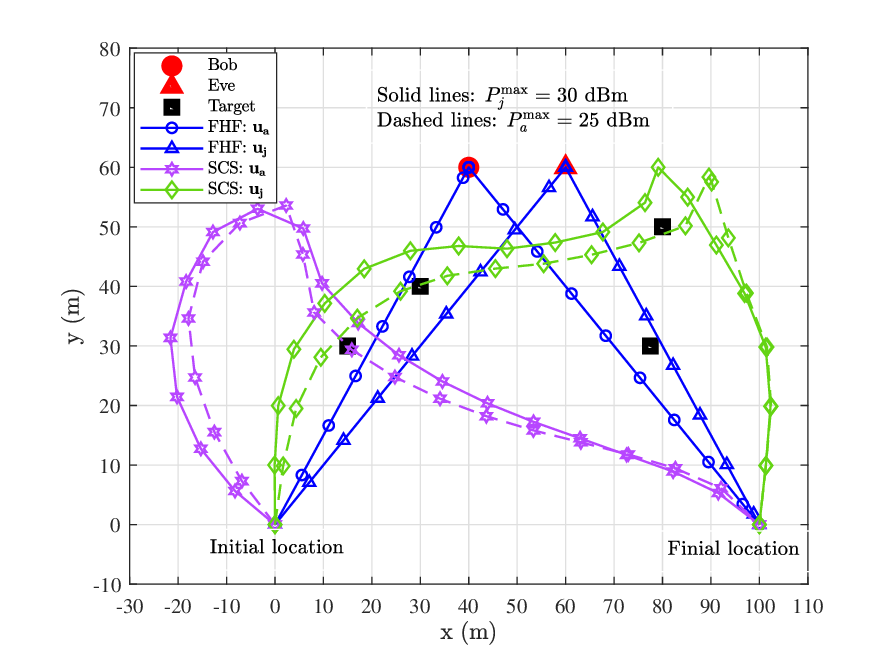}
    \caption{UAV trajectories obtained by different schemes.}
    \label{fig:trajectory}
\end{center}
\vspace{-0.2in}
\end{figure}

\begin{figure}[tt]
 \begin{center}
    \includegraphics[width=3.2in]{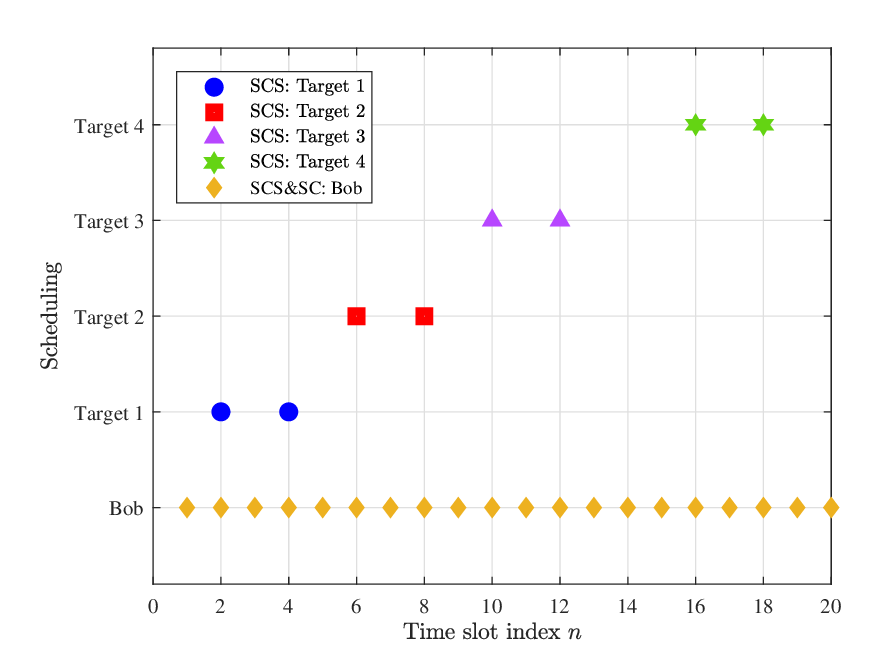}
    \caption{Selected time slot indices for target sensing.}
    \label{fig:target_slot}
\end{center}
\vspace{-0.2in}
\end{figure}

\begin{figure}[tt]
 \begin{center}
    \includegraphics[width=3.2in]{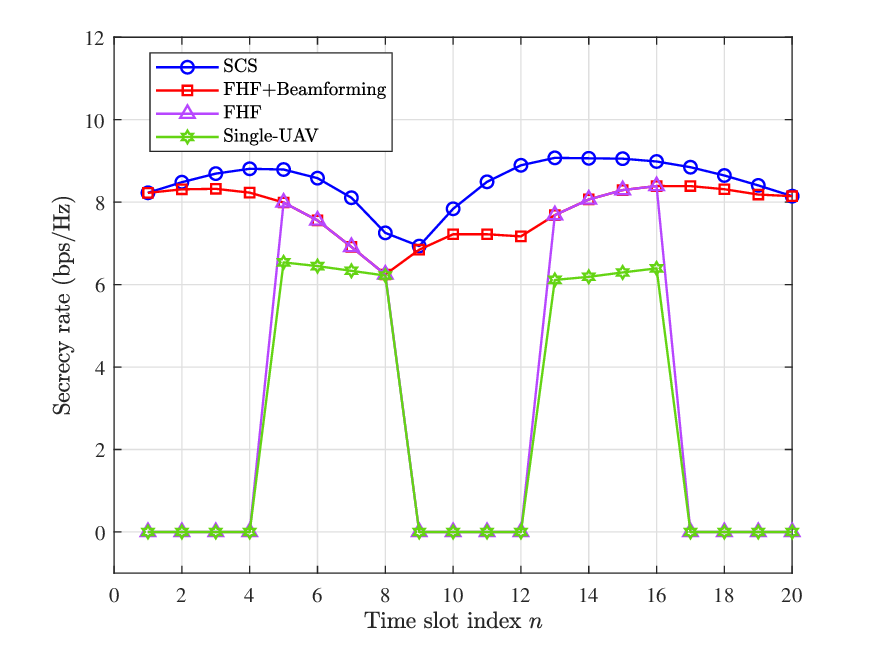}
    \caption{Secrecy rate at each time slot.}
    \label{fig:time_rate}
\end{center}
\vspace{-0.2in}
\end{figure}

Fig. \ref{fig:trajectory} illustrates the dual-UAV trajectory of the proposed SCS scheme. In Fig. \ref{fig:trajectory}, we set $P_a^{\max} = 25$ dBm, while Bob and Eve are closely located. As shown in Fig.  \ref{fig:trajectory}, both UAVs maneuver along the arc-shaped trajectories when the proposed SCS scheme is employed. Specifically, Alice flies closely toward Bob while maneuvering away from Eve to prevent data leakage. Conversely, Jack maneuvers toward Eve and flies away from Bob to suppress eavesdropping via transmitting AN. 
Compared to the trajectory of the FHF scheme, in the proposed SCS scheme, Alice and Jack avoid getting close to Eve and Bob, respectively, thus significantly decreasing the reception quality of the eavesdropping. By exploiting the dual-UAV maneuverability, 
Fig. \ref{fig:trajectory} shows that Alice and Jack collaborate to enhance the performance of secure communication.
  
Fig. \ref{fig:target_slot} presents the selected time slots for the SCS purpose, which in turn determine the dual-UAV's sensing locations based on the previously optimized trajectory. 
As clarified by the results in Fig. \ref{fig:target_slot}, the proposed greedy algorithm can effectively select the time slots for sensing each target. Moreover, Fig. \ref{fig:time_rate} gives out the secrecy rates achieved by different schemes across the time slots. As indicated by the results in Fig. \ref{fig:time_rate}, the proposed SCS scheme achieves the highest secrecy rates in all the time slots. Since the SCS scheme achieves the highest secrecy rates in all the time slots, the resulted ASR of the SCS scheme is also the highest among the benchmark schemes. However, the secrecy rates achieved by the FHF and single-UAV schemes in most of the time slots are zeros, which clarifies that the dual-UAV deployment and joint optimization of trajectory and beamforming are necessary to enhance the secure communication performance in the considered A2G-ISAC system. 

\begin{figure}[tt]
 \begin{center}    
 \includegraphics[width=3.2in]{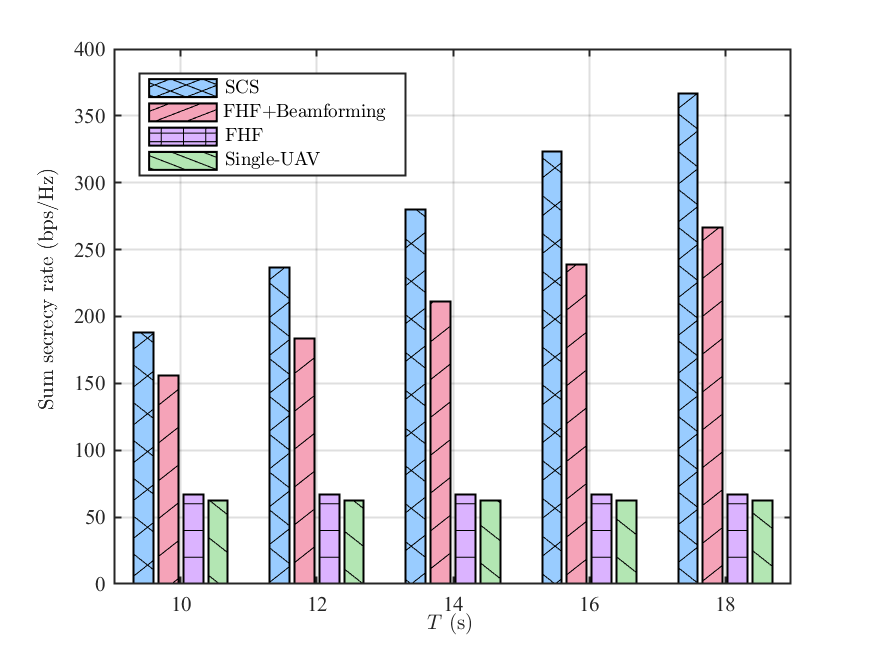}
 \caption{Sum secrecy rate versus task period duration.}
 \label{fig:T}
\end{center}
\vspace{-0.2in}
\end{figure}

Fig. \ref{fig:T} demonstrates the sum secrecy rate performance versus the task period duration $T$. The results in Fig. \ref{fig:T} show that the sum secrecy rates achieved by the SCS and FHF+Beamforming schemes increase with increasing $T$. For all the considered task durations, the proposed SCS scheme achieves the highest sum secrecy rate, while the FHF+Beamforming scheme achieves the second highest sum secrecy rate. On the other hand, the sum secrecy rates achieved by the FHF and single-UAV schemes remain the same under all the task period durations. The reason is that the adopted FHF trajectory keeps Alice and Jack hovering on top of Bob and Eve, respectively, as long as possible, which results in zero contribution to the sum secrecy rate due to the closely located Bob and Eve. However, the proposed SCS scheme keeps Alice and Jack from getting close to Eve and Bob, respectively, thus achieves a positive secrecy rate compared to the FHF trajectory.  
 
\begin{figure}[tt]
 \begin{center}
\includegraphics[width=3.2in]{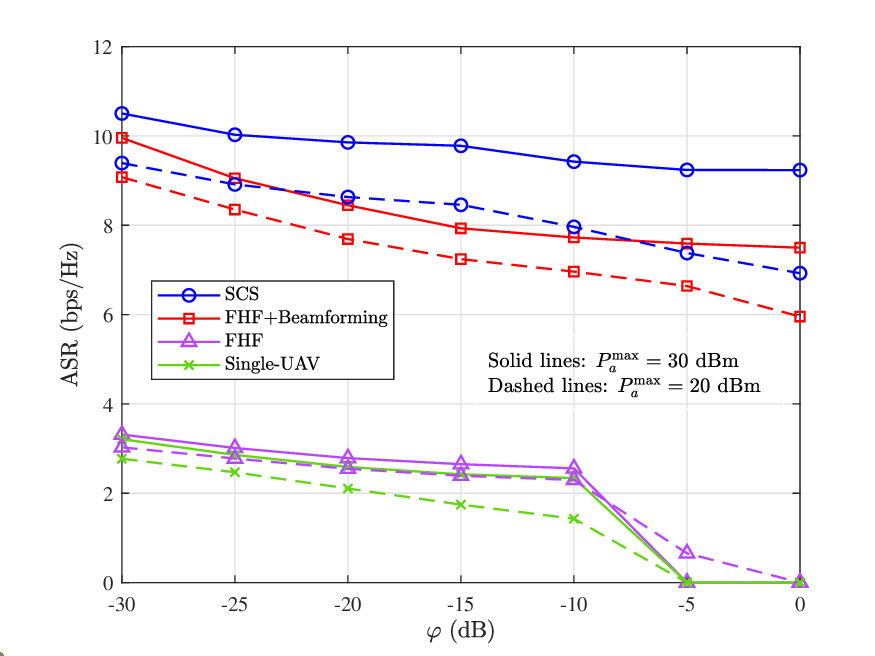}
    \caption{ASR under different values of residual inteference level.}
    \label{fig:cancellion factor}
\end{center}
\vspace{-0.2in}
\end{figure}

\begin{figure}[tt]
 \begin{center}
\includegraphics[width=3.2in]{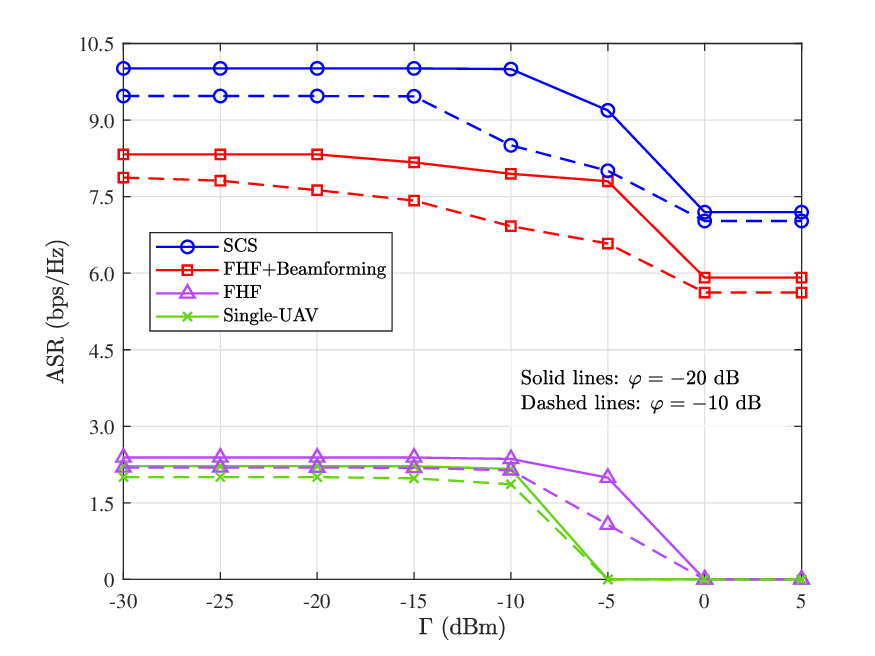}
    \caption{Trade-off between ASR and sensing performance.}
    \label{fig:sensing threshold}
\end{center}
\vspace{-0.2in}
\end{figure}

\begin{figure}[tt]
 \begin{center}
 \includegraphics[width=3.2in]{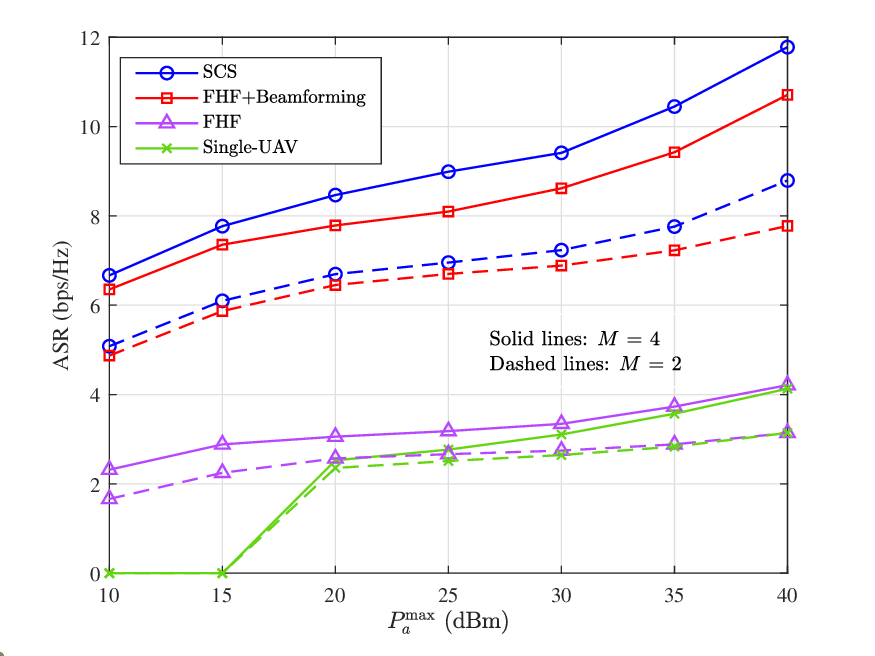}
    \caption{ASR under the different maximum transmit powers of Alice.}
    \label{fig:pa}
\end{center}
\vspace{-0.2in}
\end{figure}

\begin{figure}[tt]
 \begin{center}    \includegraphics[width=3.2in]{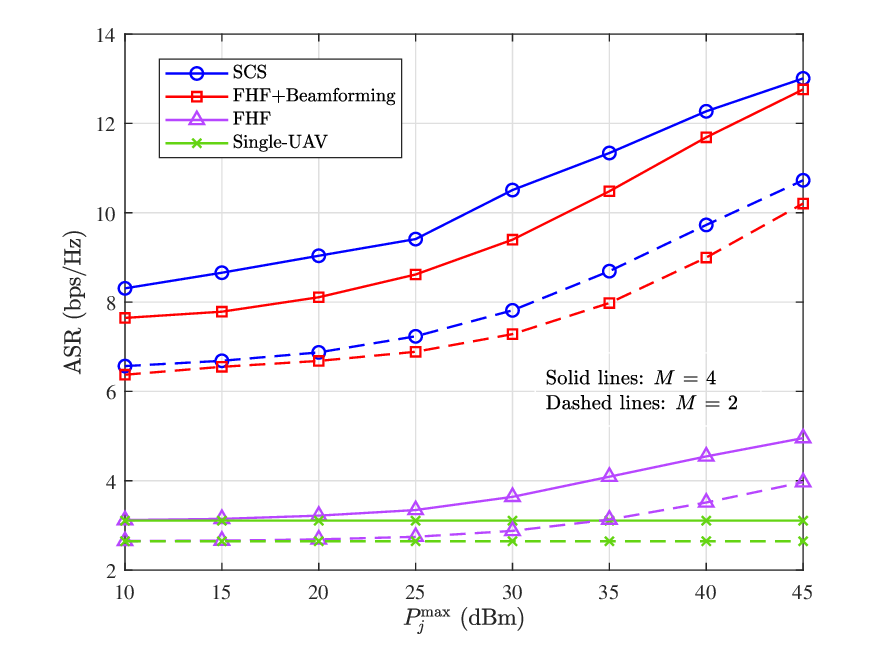}
    \caption{ASR under the different maximum transmit powers of Jack.}
    \label{fig:pj}
\end{center}
\vspace{-0.2in}
\end{figure}

Fig. \ref{fig:cancellion factor} presents the impacts of residual interference level $\varphi$ ($\varphi_{rb}=\varphi_{jb} = \varphi$) on ASR. In Fig. \ref{fig:cancellion factor}, the task period duration is set as $T=12$ s. As $\varphi$ increases, all schemes exhibit a decrease in ASR due to a weakened SIC capability, which amplifies the effects of residual interference on communication links. In the considered whole $\varphi$ region, the SCS schemes achieves the highest ASR. Moreover, the FHF and single-UAV schemes achieve a zero ASR in the high $\varphi$ region due to the high residual interference. The results in  Fig. \ref{fig:cancellion factor} further reveal that a lower ${P}_a^{\rm{max}}$ exacerbates a ASR degradation under high $\varphi$, emphasizing the critical balance between transmit power and interference management. This validates SCS scheme's adaptability in complex environments through dual-UAV coordination and maneuverable jamming. 

Fig. \ref{fig:sensing threshold} shows the trade-off between the ASR and sensing performance under a task period duration of $T=14$ s. With increasing sum-beampattern gain threshold $\Gamma$, the ASRs achieved by all the schemes decrease due to a stricter sensing performance requirement. Also, the results in Fig. \ref{fig:sensing threshold} show that the increasing in the residual interference level $\varphi$ accelerates the ASR decline. 
Among all the schemes, the proposed SCS scheme achieves the highest ASR in the whole $\Gamma$ region.  
However, in the high $\Gamma$ region, the ASR performance experiences a floor due to more resources are allocated to ensure the sensing performance at the cost of compromising the secure communication performance. Conversely, when $\Gamma$ decreases, the ASR asymptotically approaches a ceiling. This occurs because the relaxed sensing requirements enable the dual-UAV to allocate more transmit power for improving ASR. The results in Fig. \ref{fig:sensing threshold} validate that the proposed SCS scheme can well balance the trade-off between the ASR and sensing performance.  

Fig. \ref{fig:pa} investigates the impacts of Alice's maximum transmit power $P_{a}^{\max}$ on ASR. The experimental results indicate that the ASR  increases with increasing $P_a^{\rm{max}}$. Furthermore, when $P_a^{\rm{max}}$ is too low, even the introduction of the jamming UAV Jack does not significantly improve the ASR. This suggests that under low transmit power conditions, the introduction of a jamming UAV has a limited contribution on enhancing the secure communication performance. Further observation demonstrates that the proposed SCS scheme achieves a significantly higher ASR as $P_{a}^{\max}$ increases, with its growth rate exceeding other schemes, while using $M = 4$ antennas consistently outperform $M = 2$. In addition, the ASR of the single-UAV scheme approaches zero in the low transmit power range due to resource limitations, which highlights the role of multi-UAV collaboration in improving the efficiency of the utilization of transmit power and trajectory. By optimizing UAV trajectory and beamforming, the proposed SCS scheme can efficiently enhance the secure communication performance.

Fig. \ref{fig:pj} presents the impacts of the maximum jamming power $P_{j}^{\max}$ on ASR. The experiment results show that the ASR of all schemes increases with increasing $P_{j}^{\max}$, while the proposed SCS scheme maintains a clear advantage in the whole $P_{j}^{\max}$ range. When $M = 4$ antennas are adopted, the ASR performance is further enhanced compared to $M = 2$, indicating that increasing the number of the transmit antennas can effectively jamming the eavesdropping link. Although increasing $P_{j}^{\max}$ brings benefits to all schemes, the ASR improvement of the FHF scheme is significantly lower than that of the proposed SCS scheme due to the lack of the joint optimization of the trajectory and beamforming. Moreover, the ASR achieved by the single-UAV scheme remains constant in the whole $P_{j}^{\max}$ range due to the lacking of the UAV cooperation and beamforming optimization.  

\begin{figure}[tt]
 \begin{center}
\includegraphics[width=3.2in]{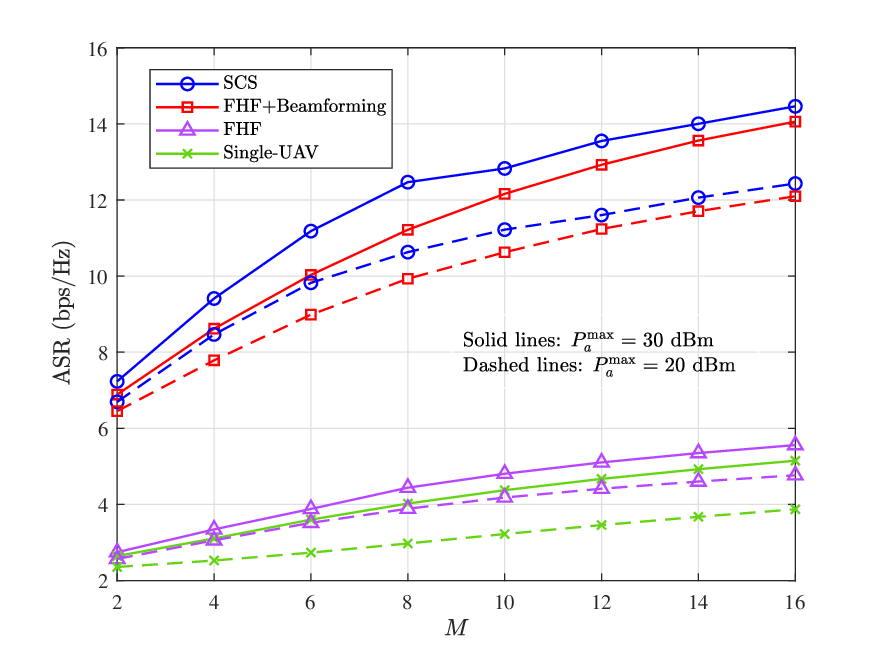}
    \caption{ASR under the different numbers of 
transmit antennas.}
    \label{fig:antenna}
\end{center}
\vspace{-0.2in}
\end{figure}

Fig. \ref{fig:antenna} illustrates the relationship between ASR and the number of transmit antennas, $M$. 
With increasing $M$, the ASRs achieved by all the schemes increase, while the proposed SCS scheme achieves the highest ASR. When $P_a^{\max}$ increases from 20 dBm to 30 dBm, the achieved ASR of all the schemes also increase. In addition, the FHF+Beamforming scheme exhibits a slower ASR growth than the proposed SCS scheme, while the FHF and single-UAV schemes achieve a quite small ASR in the considered $M$ range. The results in Fig. \ref{fig:antenna} clarify that the beamforming is pivotal to enhance system performance. Besides, the simulation results also verify that by jointly optimizing the dual-UAV trajectory and dual-UAV beamforming, the secure communication and target sensing performance can be significantly enhanced by the proposed SCS scheme for the considered A2G-ISAC system.  

\section{Conclusions}

In this work, we have proposed the dual-UAV-enabled SCS scheme for the A2G-ISAC system to enhance both the secure communication and target sensing performance. For the first time, the source and jamming UAVs construct a hybrid monostatic-bistatic radar to sense multiple ground targets, while the jamming UAV is deployed to further improve the secure communication performance. To maximize the ASR for the considered A2G-ISAC system, the dual-UAV trajectory and dual-UAV beamforming have been tackled for the SC and SCS purposes, sequentially. For the SC purpose, we have proposed a BCD algorithm to optimize the dual-UAV trajectory and dual-UAV beamforming by using the SDR and trust-region SCA approaches. For the SCS purpose, we have proposed a greedy algorithm to determine the dual-UAV sensing locations and a penalized beamforming optimization. Simulation results have verified that the proposed SCS scheme achieves the highest ASR performance over the existing benchmark schemes. The sensing performance has been effectively improved by exploiting the jamming UAV with the optimized trajectory and beamforming. Also, simulation results have revealed the impacts of residual interference level on the secure communication and target sensing performance. 

\begin{balance}
\bibliography{ref}
\end{balance}

\end{document}